\documentclass[referee]{aa} 
\usepackage{graphicx}
\usepackage{txfonts}
\usepackage{epstopdf}
\usepackage{lineno}
\usepackage{amstext}

\begin{document}


   \title{Radial velocity map of solar wind transients in the field of view of STEREO/HI1 on 3 and 4 April 2010}
   \titlerunning{Radial velocity map of solar wind transients}

   \author{Xiaolei Li
          \inst{1,2}
          , Yuming Wang
          \inst{1,2,3}\fnmsep\thanks{Corresponding author}
          , Jingnan Guo
          \inst{1,2}
          , Rui Liu
          \inst{1,2,3}
          ,
          \and
          Bin Zhuang
          \inst{4}
          }
   \authorrunning{Li et al.}
   \institute{CAS Key Laboratory of Geospace Environment, Department of Geophysics and Planetary Sciences, University of Science and Technology of China,Hefei 230026, China\\
         \email{lxllxl@mail.ustc.edu.cn, ymwang@ustc.edu.cn}
         \and
             CAS Center for Excellence in Comparative Planetology, University of Science and Technology of China, Hefei 230026, China
        \and
             Mengcheng National Geophysical Observatory, University of Science and Technology of China, Mengcheng 233527, China
        \and
             University of New Hampshire, Durham, New Hampshire, United States
             }


  \abstract
   {The solar wind transients propagating out in the inner heliosphere can be observed in white-light images from Heliospheric Imager-1 (HI1), an instrument of the Sun Earth Connection Coronal and Heliospheric Investigation (SECCHI) on board the Solar Terrestrial Relations Observatory (STEREO), from two perspectives. The spatial velocity distribution inside solar wind transients is key to understanding their dynamic evolution processes.}
   {We generated a velocity map of transients in 3D space based on 2D white-light images and used it to estimate the expansion rate as well as some kinematic properties of solar wind transients.}
  {Based on the recently developed correlation-aided reconstruction (CORAR) method in our previous work, which can recognize and locate 3D solar wind transients from STEREO/HI1 image data, we further developped a new technique for deriving the spatial distribution of the radial velocities of the most pronounced features inside solar wind transients.}
   {The technique was applied to events including a coronal mass ejection (CME) and three small-scale transients, so-called blobs, observed by HI1 on 3-4 April 2010 to reconstruct their radial velocity maps. The results match the forward-modeling results, simulations, and in situ observations at $1$ AU fairly well. According to the obtained spatial distributions of height and radial velocity of the CME, we analyzed the self-similarity of the radial expansion of the CME ejecta. The dimensionless radial expansion rate of the northern and middle parts of the CME ejecta varies in the range of  $0.7 - 1.0$ at heliocentric distance between $25 R_{\odot}$ and $55R_{\odot}$ and the rate of the southern part in the range of  $0.3 - 0.5$, suggesting that the CME structure was distorted and shaped by the ambient solar wind. The technique we developed is expected to be applied to more events.}
  {}

   \keywords{methods: data analysis - Sun: Coronal mass ejections (CMEs) - solar wind}

   \maketitle

\section{Introduction}

   Solar wind transients are macroscopic inhomogeneous structures in the solar wind originating from the solar corona. Coronal mass ejections (CMEs) are large-scale solar wind transients, typically consisting of ejecta, a driven shock, and a shock sheath in between \citep[e.g.,][]{2012LRSP....9....3W} if they are fast enough. The ejecta is believed to be a huge bending flux rope that erupts from the solar surface \citep[e.g., ][]{1997ApJ...490L.191C, 2004A&A...422..307C, 2006ApJ...652..763T}. In white-light images, the ejecta structure usually shows a bright front, which is also called the leading edge, a dark cavity, and a bright core, if any \citep{2006cme..book..127S}. In addition to CMEs, high-density small-scale solar wind transients called "blobs" \citep{1997Measurements,2017ApJ...851...32S} propagate from the near boundaries of small coronal holes or the top of the streamers into the heliosphere with the slow solar wind. They are therefore thought to be good tracers of the slow solar wind in observations \citep{L2018Deprojected}. A wide range of undefined small-scale solar wind transients or disturbances have also been recorded in in situ and remote-sensing observations \citep[e.g.,][]{2009SoPh..256..327K, 2011ApJ...734....7R, 2018ApJ...862...18D}. These solar wind transients and the continuum background solar wind constitute the plasma environment that surrounds our planet. The velocity distribution inside solar wind transients is a key information for us to understand their evolution in the heliosphere and the effects on space weather.

   Although some spacecraft, such as Wind around the first Lagrange point of the Sun-Earth system since 2004 \citep{1995SWE}, provide in situ measurements of the solar wind velocity with high temporal resolution and high accuracy, the number and coverage of spacecraft in the heliosphere is very limited, and therefore it is impossible to obtain the velocity distribution of transients even in two or more spatial dimensions from in situ observations. Interplanetary scintillation (IPS) measurements through many radio sources can detect the density variation of transients and reconstruct their 3D velocity map in the whole inner heliosphere with the computer-assisted tomography (CAT) technique \citep{Jackson1998Heliospheric, 2010Ooty}, for instance, in relatively low temporal and spatial resolution.

   The density variation of transients leads to an intensity variation in Thomson-scattered light, which can also be remotely observed in relatively high temporal and spatial resolution by space-borne white-light imagers such as the Large Angle and Spectrometric Coronagraph (LASCO) \citep{1995The} on board the Solar and Heliospheric Observatory (SOHO), the coronagraphs (COR1 and COR2) and heliospheric imagers (HI1 and HI2) in the Sun Earth Connection Coronal and Heliospheric Investigation (SECCHI) suite \citep{2008SSRv..136...67H} on board the Solar Terrestrial Relations Observatory (STEREO), Wide-field Imager for Solar PRobe (WISPR) \citep{2016SSRv..204...83V} on board Parker Solar Probe (PSP) \citep{2016SSRv..204...83V} and SoloHI \citep{howard2019solar} on board Solar Orbiter \citep{2020A&A...642A...1M}. However, due to the lack of the information of the depth along the line of sight (LOS), a 2D image cannot directly convey the complete 3D information of solar wind transients. In practice, the plane of sky (POS, i.e., the plane perpendicular to the observer-Sun line and crossing the solar center) is usually chosen as a projection plane for solar wind transients along the LOS, especially for a coronagraph, to study their kinematics. Alternatively, \citet{2006ApJ...642.1216V} proposed that a Thomson sphere, which is a sphere with a diameter of the observer-Sun line, should replace the POS. However, \citet{2012ApJ...752..130H} suggested that there is no significant difference in brightness for a solar wind transient on or near the Thomson sphere. An accurate derivation of the 3D position and velocity of transients still relies on other methods.

   The leading edges of the transients normally leave bright traces in the images of visible light, inspiring many methods that were developed to derive their locations and velocities, such as the icecream cone model \citep{1984ApJ...280..428F}, the graduated cylindrical shell (GCS) model \citep{thernisien2011implementation}, geometric triangulation methods \citep{2010ApJ...710L..82L}, mask-fitting methods \citep{feng2012morphological}, and trace-fitting methods including the point-p, fixed-$\Phi$, harmonic mean, and self-similar expansion fitting methods \citep[e.g., ][]{sheeley1999continuous, 2006JGRA..111.4105H, 2012ApJ...750...23D, mostl2013speeds}. To derive the velocity distribution inside one transient rather than only at its leading edge, some other techniques have been proposed. \citet{colaninno2006analysis} applied an optical flow tool to extract the velocity vector of a coronal mass ejection (CME) in digital images. \citet{feng2015radial} derived the radial velocity profiles of the whole CME from the spatial distribution of its density given by the mass continuum equation. A cross-correlation method was applied to derive continuous 2D speed maps of a CME from coronagraphic images by \citet{bemporad2018measuring}. In their work, the radial shift pixel by pixel is determined by maximizing the cross correlation between the signal in a radial window at one frame and the signal in a radial shifted window at the previous frame, and the radial speed just equals the radial shift over the time interval between the two frames. \citet{ying2019first} improved this cross-correlation method by analyzing data in three steps: forward step (FS), backward step (BS), and average step (AS). In the FS (BS), the 2D velocity map between the current and the previous (next) frame is constructed with almost the same method as \citet{bemporad2018measuring}. In the AS the average, velocity is obtained from the FS and BS. The velocities derived by all these methods are the component of the flow velocity vector projected onto the POS. This may underestimate the velocity especially for transients that do not propagate in the POS. Methods such as the polarizaition ratio technique \citep{moran2004three, deforest20173d} or the local correlation tracking (LCT) method \citep{mierla20093d} can derive the 3D geometric information of the whole transients, but not the velocity distribution. \citet{bemporad2018measuring} chose the propagating direction averaged over the whole CME derived by the polarization ratio technique to correct the radial speed in the 2D maps, but the key information along the LOS is still lacking.

   \citet{li2018reconstructing}, hereafter referred to as Paper I, identified the transients near the Sun-Earth line using a correlation analysis of STEREO/HI1 images from two perspectives. \citet{2020Reconstructing}, hereafter Paper II, further recognized and located 3D solar wind transients in the common field of view (FOV) of two HI1 imagers from two perspectives, and called this technique the correlation-aided reconstruction (CORAR) method. The reconstruction performance of CORAR method is tested to be fairly well when the separation angle of the twin STEREO spacecraft is between about $120^\circ$ and $150^\circ$ \citep{2020AdSpR..66.2251L,2021arXiv210103276L}.

   In this paper, we further reconstruct the radial velocity map of transients in 3D space with a technique called maximum correlation-coefficient localization and cross-correlation tracking (MCT). In Section 2 we introduce the data and the events, including one CME and three blobs, to which we apply this new technique. The description of the MCT technique and its application results are presented in Section 3 and 4. In Section 5 we analyze the expansion of the CME ejecta between a heliocentric distance of $25$ and $55 R_{\odot}$. Discussion and summary are given in Section 6.

\section{Data and events}

   \subsection{Data}

   Similar to Paper II, we used the white-light images data taken by the STEREO/HI1 cameras on 3 and 4 April 2010 to demonstrate the MCT technique. The separation angle of STEREO was $140^\circ$ then (Fig.1f). STEREO HI1 images have a wide FOV of $20^\circ$ by $20^\circ$ centered at $14^\circ$ in elongation near the ecliptic plane with a pixel resolution of $1024\times1024$ and a cadence of $40$ minutes \citep{harrison2008first}. To show the solar wind transients clearly, as what we did in Paper I and II, we processed the calibrated STEREO/HI1 Level 2 images \citep{eyles2009heliospheric} with$\text{}\text{}$ one-day background emissions, defined as the average of the lowest $25\%$ of the data in a running window of $\text{one}$ day on a pixel-by-pixel basis with a shifted running difference and $3\times3$ median filtering.

   \subsection{Events}

\begin{figure*}[htbp]
    \begin{center}
      \includegraphics[width=\hsize]{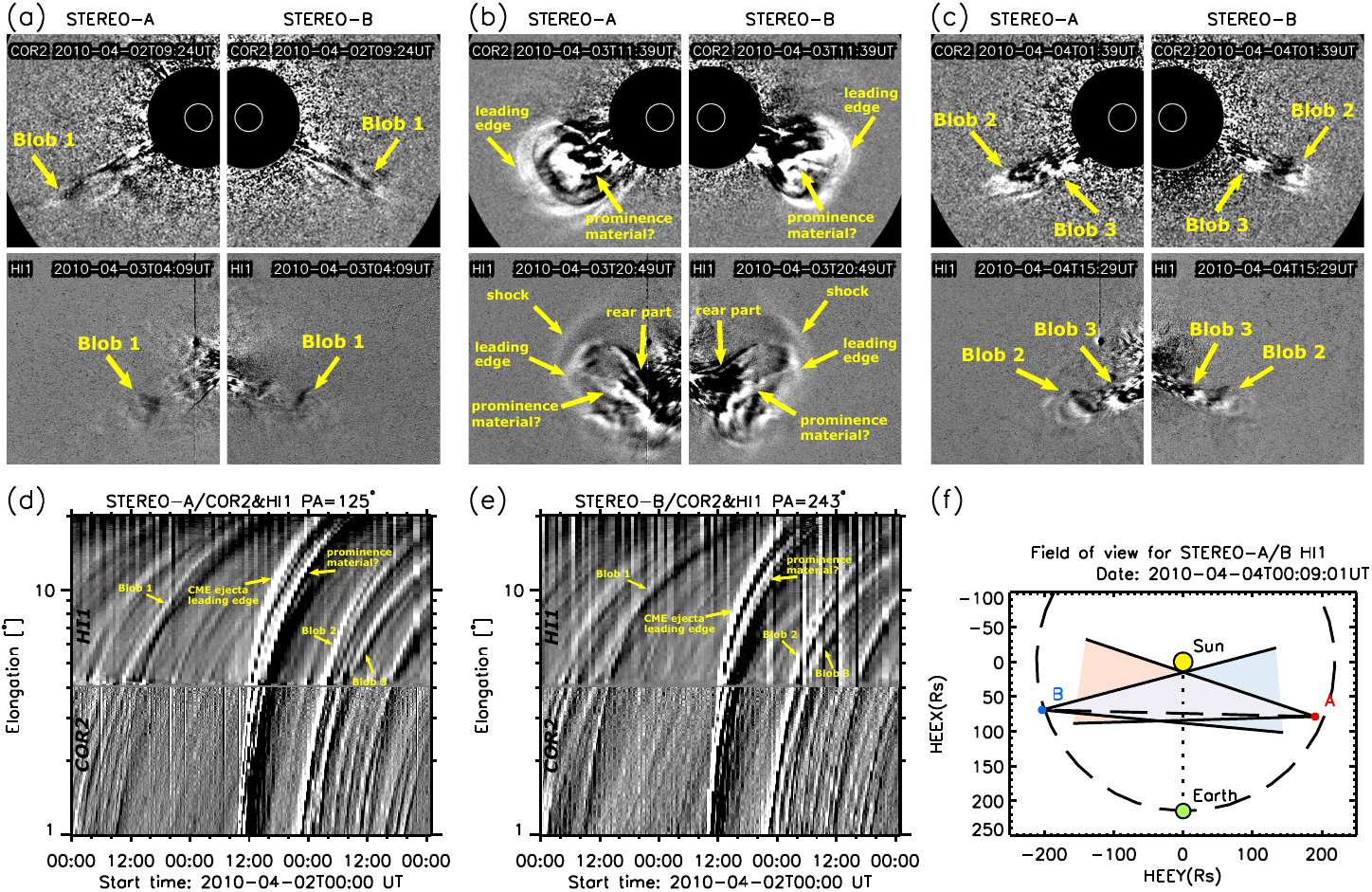}
    \end{center}
      \caption{Main solar wind transients, including three blobs and one CME, observed from the perspective of STEREO-A/B on 3 and 4 April 2010. Panels (a)-(c) show them in the running-difference images of COR2 and HI1 on board STEREO-A/B. Panel (d)/(e) shows the time-elongation plot along the PA of $125^\circ$/ $243^\circ$ for STEREO-A/B COR2 and HI1. Blobs 1-3 and the shock, the ejecta leading edge, the potential prominence material, and the ejecta rear part of the CME are marked in these panels. Panel (g) displays the position of the Sun, Earth, and STEREO-A/B on the ecliptic plane at 00:09:01 UT on 4 April 2010. The region between the two solid black rays from STEREO-A or STEREO-B indicates the FOV of HI1, and the lilac region represents the common FOV of STEREO-A and B HI1s.
              }
         \label{Fig0}
   \end{figure*}

   On 3 and 4 April 2010, three slow blob-like small-scale transients, the blobs, labeled Blob 1-3, one fast CME, and its driven shock were observed by the two STEREO/HI1 cameras (see the bottom row of Fig. \ref{Fig0}a-\ref{Fig0}c, see also Section 4.1 in Paper II for more details about the identification of these transients) and could be traced back to at least the FOVs of the coronagraphs (see the top row of Fig. \ref{Fig0}a-\ref{Fig0}c). Especially in the time-elongation plots along the position angle (PA) of $125^\circ$ for STEREO A and $243^\circ$ for STEREO B, the leading edges of these transients could be clearly traced from COR2 to HI1 (Fig. \ref{Fig0}d and \ref{Fig0}e). The CME structure is complicated in HI1 running-difference images (Fig. \ref{Fig0}b). In the COR2 FOV, it has a clear bright leading edge and a bright core. The shock is hard to be distinguished from the bright leading edge in the COR2 FOV. When the CME propagated into the HI1 FOV, however, the shock standing-off distance increased because the CME ejecta decelerated, and therefore the shock became distinguishable (Fig.1b). The flattening of the CME ejecta can be revealed in HI1 images compared with the COR2 images. A bright ribbon-like structure extending latitudinally between the leading edge and the rear part of the ejecta corresponds to the bright core in COR2 images (Fig.1d and 1e) and was suggested to be potential prominence material in \cite{2011ApJ...735....7R}. The shock front, shock sheath, and the following CME ejecta were observed by in situ instruments at 1 AU \citep{2010GeoRL..3724103M, 2011ApJ...735....7R} (see Fig. \ref{Fig6}). The two-day imaging data contain both large- and small-scale fast and slow solar wind transients, and the STEREO spacecraft separated by an ideal angle. We therefore selected these events to demonstrate our velocity reconstruction technique.

\section{3D probable position maps of transients: Maximum correlation-coefficient localization}

   \subsection{Procedures: CORAR method and maximum cc localization}

   In Paper II, the CORAR method \citep[][Paper II]{2020Reconstructing} was developed and applied to the HI1 data during 3-4 April 2010 to identify, locate, and trace the solar wind transients in 3D space. Its main procedures are (1) choosing any radial line with one end at the solar center as the baseline and projecting two HI1 images onto the meridian plane where the baseline lies; (2) for transients on the baseline, the local projected images are similar around the transients, resulting in a high value of Pearson correlation coefficient (cc), while for transients away from the baseline, the local projected images should be dissimilar to each other, resulting in a low cc value; (3) adjusting the latitude or longitude of the baseline at a step of $1^{\circ}$ in 3D space and repeating steps (1) and (2), a 3D cc map can be generated. A higher value of cc represents a more probable location of a transient feature, and the high-cc regions (defined by $cc\geq0.5$, see Section 2 of Paper II for details) are regarded as the 3D reconstructed solar wind transients.

\begin{figure*}[htbp]
    \begin{center}
      \includegraphics[width=\hsize]{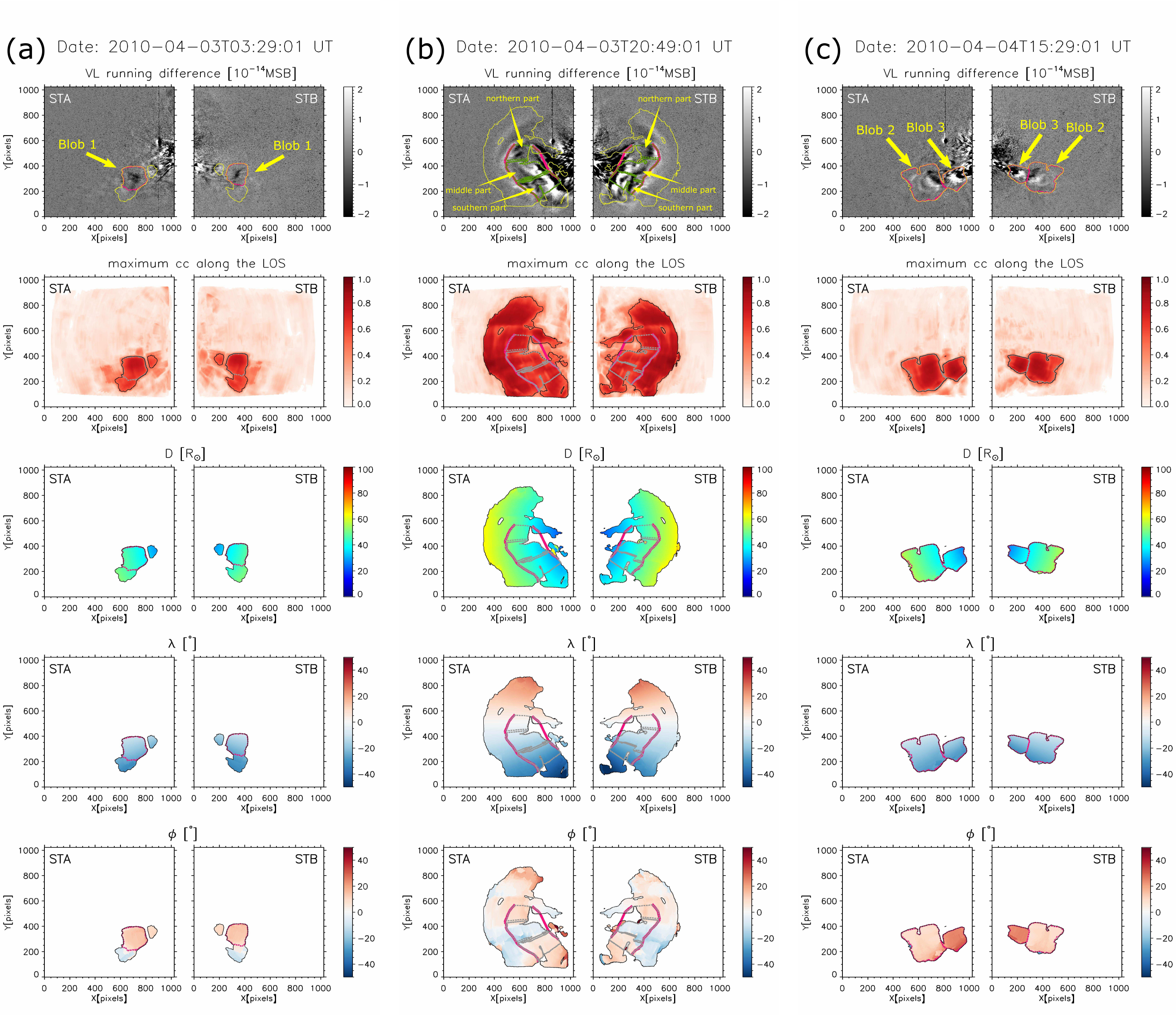}
    \end{center}
      \caption{3D pp map of the transients from the perspective of STEREO-A/B. Panel (a)-(c) display Blob 1 at 03:29 UT on 3 April 2010, the CME at 20:49 UT on 3 April 2010, and Blobs 2 and 3 at 15:29 UT on 4 April 2010, respectively. The left and right columns in each panel correspond to the results from the STEREO-A and STEREO-B. The five rows from the top to the bottom show the white-light (i.e., visible light (VL)) running difference images (in units of mean solar brightness (MSB)), maximum cc along the LOS, the distance from the solar center ($D$), the latitude ($\lambda$), and the longitude ($\phi$) of the solar wind transients in HEE coordinates. The solid yellow lines in the top row and the solid black lines in the other rows denote the high-cc regions (see main text for details). In Panels (a) and (c) from the top to the bottom, the blob-like transients are circled with dashed purple curves in rows 1 and 3-5 and with dashed gray curves in row 2. In Panel (b) the leading edge and rear part of the ejecta are marked with thick solid purple curves, and the northern part ($-5^{\circ}<\lambda<5^{\circ}$), middle part ($-20^{\circ}<\lambda<-5^{\circ}$), and southern part ($-35^{\circ}<\lambda<-20^{\circ}$) of the ejecta between the leading edge and the rear part are circled with dashed green curves in row 1 and dashed gray curves in rows 2-5 from the top to the bottom.
              }
         \label{Fig1}
   \end{figure*}

   Based on the 3D cc map, here we inspect the cc distribution along the LOS for each pixel in the corresponding HI1 image and treat the location of the highest cc value as the most probable position (pp) of the transient feature at this pixel. We display the highest LOS cc values for all the pixels within the recognized high-cc regions enclosed by black lines in the second row of Figure \ref{Fig1} or \ref{Fig2} for STEREO-A/B. The 3D position information of these highest cc is then derived, including the radial distance ($D$), latitude ($\lambda$), and longitude ($\phi$) in the heliocentric Earth ecliptic (HEE) coordinates (shown in rows 3-5 of Figure \ref{Fig1}, an animation, M1.mp4, can be found in the supporting materials). We call them 3D pp maps of transients from one perspective. The distance and latitude of the pp are obvious in the 2D HI1 images, but the longitude is unique information provided by our technique. The pp map directly displays the probable 3D locations and the radial propagating directions of all the high-cc features in the transients.

   \subsection{Analysis of the 3D pp maps of the transients on 3-4 April 2010}

   For the blobs, we calculated the mean values and standard deviations of their 3D pp, and display them in rows 2-4 in Figure \ref{Fig3} as shown by the red and blue crosses and error bars. The averaged 3D pp from STEREO-A and B agree well with each other and are almost at the center and within the uncertainty range of the icecream cone model results (see Section 4.4 of Paper II for details), which are displayed as black lines and shadow regions.

   For the CME, we had compared the results of CORAR method with those from other forward-modeling methods including the GCS model \citep{thernisien2011implementation}, the KS06 model \citep{2006ApJ...652.1740K, 2012JGRA..117.4105X}, and the empirical reconstruction model \citep{2011ApJ...729...70W} for this CME in Paper II and found that they match well. According to the pp map, the longitudes of the CME features are found to be positive in the northern part ($-5^\circ<\lambda<5^\circ$), negative in the middle part ($-20^\circ<\lambda<-5^\circ$), and they returning to positive in the southern part ($-35^\circ<\lambda<-20^\circ$) for both STEREO-A/B HI1 images, as shown in row 5 of Figure \ref{Fig1}. This is consistent with the angular distortion of the CME revealed in Fig. 9b of Paper II. In Table \ref{Tab1} we also compare the derived probable longitude ($\Phi$) of the CME ejecta leading edge at latitude ($\lambda$) of $0{^\circ}$ and $-3{^\circ}$ in the HEE coordinates with the results of other nonforward-modeling methods. The deviation generally remains within $15{^\circ}$ for most methods, except for the harmonic mean fitting method and the self-similar expansion method \citep{2013ApJ...777..167D}, which can be attributed to the large estimation error of the two methods for high-speed transients \citep{2014ApJ...787..119M}, just as we discussed in Paper II.
With the information of the radial propagating direction of the features inside solar wind transients, the pp map was used to derive the radial velocity of the transients pixel by pixel in the next step.

\section{Radial velocity maps of transients: Cross-correlation tracking}

   \subsection{Procedures: Cross-correlation tracking}

   We go through all the pixels of interest on the pp map at a certain time. For each pixel, we know its position in the 3D space and can set a meridian plan on which the pixel lies. Then we project the HI1 images at the selected time and the adjacent (either previous or next) time on the meridian plane. By calculating the cross-correlation coefficient of the data in a sampling box around the position of the pixel on these projected images, the radial shift at which the cross-correlation coefficient reaches maximum, and therefore the radial velocity of this pixel, can be obtained. The sampling box is set to be 11-pixel (i.e., $\pm5^\circ$) wide in latitude and 41-pixel (i.e., ~$8 R_{\odot}$) long in the radial direction, and runs along the radial direction by a step of $0.2 R_{\odot}$  to search the highest cross-correlation coefficient. Similar to \citet{ying2019first}, we applied these steps to the current and previous images to derive the backward radial velocity ($v_{rB}$), and to the current and next images to obtain the forward radial velocity ($v_{rF}$). Their average value was assigned as the final radial velocity of the pixel. After going through all the pixels on the 3D pp map, the radial velocity maps can be derived (see rows 3-5 of Figure \ref{Fig2}, an animation, M2.mp4, can be found in the supporting materials).

   \subsection{Error analysis}

    The measuring error is a directly result from the error of the radial shift ($D$), $\delta D$, which is half of its step size (about $0.1 R_{\odot}$). Considering the cadence of the HI1 images, $t$, of 40 minutes, we derived the velocity error from radial shift error of
    \begin{eqnarray}
      \delta v_r=\delta (\frac{D}{t})= \frac{\delta D}{t}=\frac{0.1 R_{\odot}}{40min}= 29 km/s.
    \end{eqnarray}
    We also considered the different-apparent-leading-edge (DALE) effect \citep{liewer2011stereoscopic}, which is an error source that arises because different parts of the leading edge of the same transient are seen from two perspectives in the reconstruction with the triangulation method. The CORAR method is essentially a kind of triangulation location method: its technique for identifying the same part of the transient by maximizing the correlation coefficient instead of doing this with manual work distinguishes it from other methods. \citet{liewer2011stereoscopic} analyzed the velocity error from the DALE effect with a simple model of the leading edge as a hemispherical shell of radius $a$ with a sphere-center distance $R$ from the Sun. Assuming the leading edge located in the ecliptic plane propagates at an angle $\beta$ from the Sun-Earth line and the camera rays are considered parallel for two spacecraft at a similar angular distance $\theta$ from the Sun-Earth line but on different sides, \citet{liewer2011stereoscopic} derived the fractional error in velocity when $\theta>\beta$ for STEREO/COR images as
    \begin{eqnarray}
      \frac{\delta v_r}{v_r}=\frac{\sqrt{1+2 \frac{a \cos{\beta}}{R \sin{\theta}} +{(\frac{a}{R})}^2 {(\frac{1}{\sin{\theta}})}^2}}{1+\frac{a}{R}}-1.
    \end{eqnarray}
    For HI1 images with a large elongation angle, $\theta$ here should be replaced by the average angle between observer-object line and the Sun-Earth line. For $60^{\circ}\leq \theta \leq 90^{\circ}$, this fractional error increases as $\theta$ decreases or $\beta$ decreases and is up to $5\%$ when $a/R=0.5$ and $8\%$ when $a/R=1$. For the CME event on 3 and 4 April 2010, $\theta>75^{\circ}$, $\delta v_r/v_r$ is no more than $1.2\%$ when $a/R=0.5$ and $1.8\%$ when $a/R=1$. Even for a high speed of $1000$ km/s, the error from the DALE effect is $18$ km/s at most, which is less than that from the radial shift error. The direct error of the derived radial velocity is therefore smaller than 50 km/s.

   \subsection{Analysis of the radial velocity maps of the blob-like small transients on 3-4 April 2010}

\begin{figure*}[htbp]
    \begin{center}
      \includegraphics[width=\hsize]{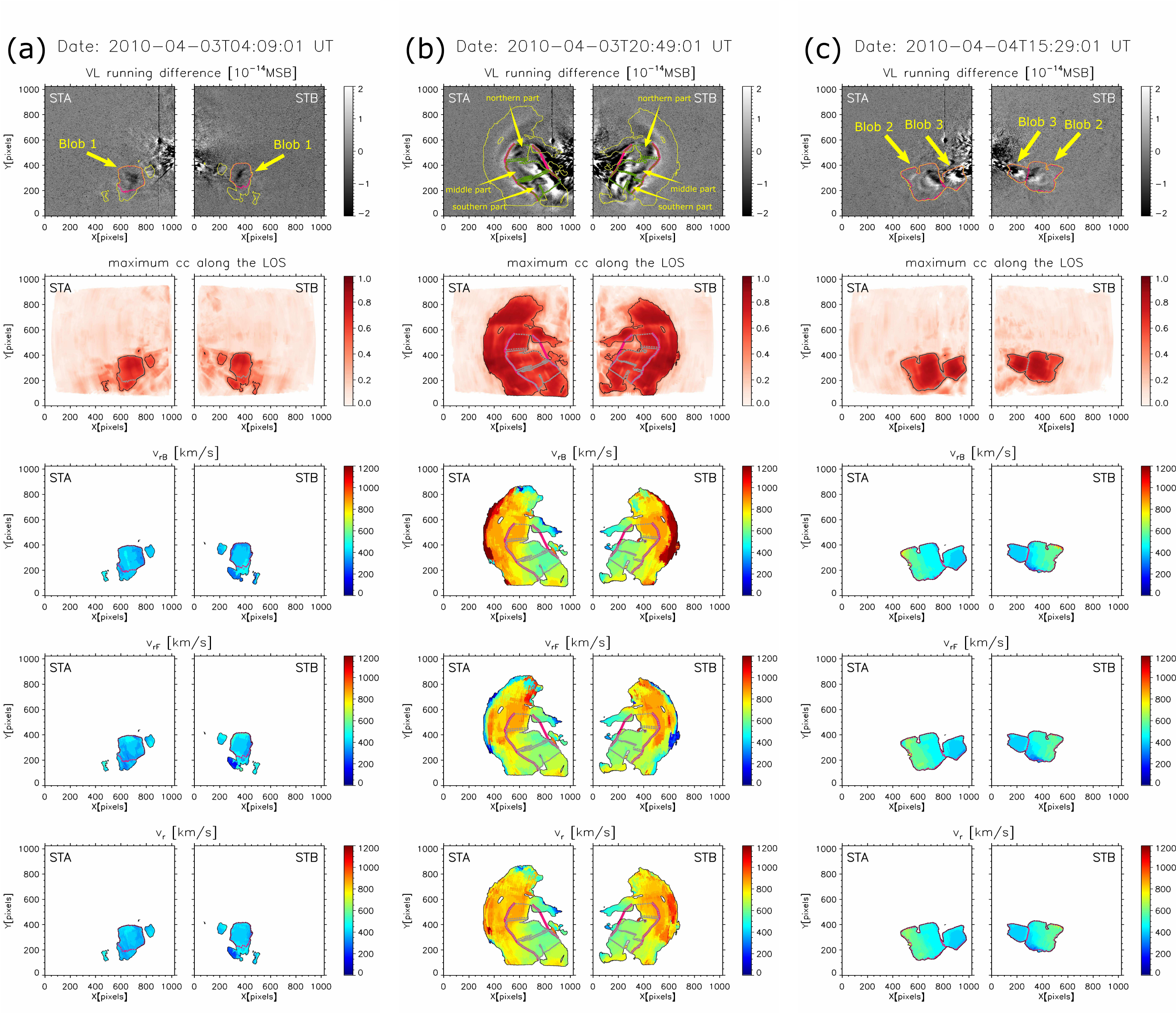}
    \end{center}
      \caption{Radial velocity map of the transients from the perspective of STEREO-A/B. From top to bottom, each panel shows the VL running difference images, maximum cc along the LOS, $v_{rB}$, $v_{rF}$ and the final radial velocity ($v_r=(v_{rB}+v_{rF})/2$). The others are same as Figure \ref{Fig1}.
              }
         \label{Fig2}
   \end{figure*}

\begin{figure*}[htbp]
    \begin{center}
      \includegraphics[width=\hsize]{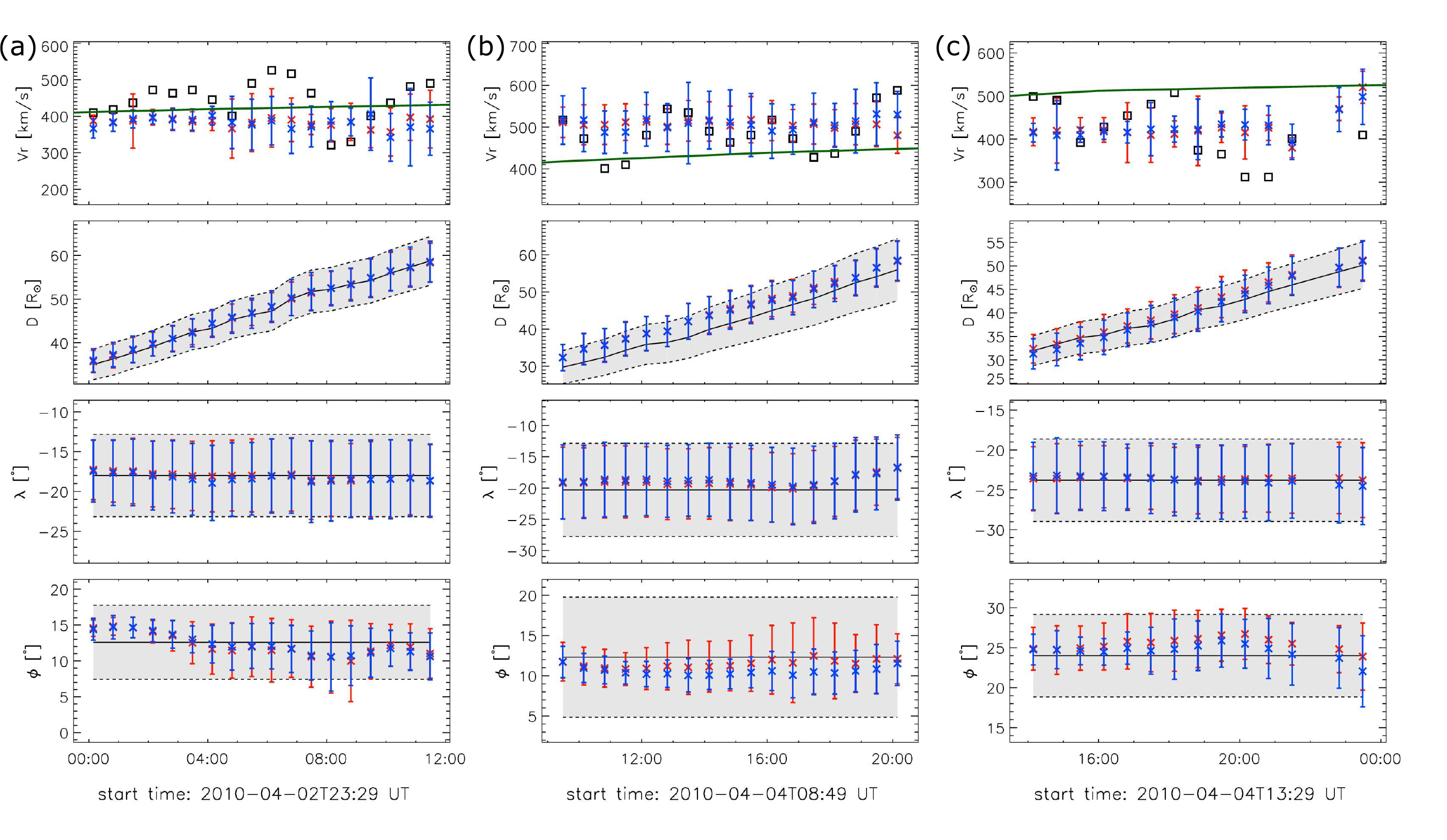}
    \end{center}
      \caption{Time-dependent radial velocity and 3D position of Blobs 1-3 in columns (a)-(c). The four rows from top to bottom show the radial velocity ($v_r$), the distance from the solar center ($D$), and the latitude ($\lambda$) and longitude ($\phi$) of these blobs in HEE coordinates. The orange and blue crosses and error bars represent the mean value and standard deviation of the velocity or 3D pp of the blobs in the FOV of STEREO-A/B. In the top row, the thick solid green lines display the simulated solar wind radial velocity with the ENLIL code, and the black squares are the results of icecream cone model. In the other rows, the black lines and the shadow regions represent the center and the range of the fitting icecream ball.
              }
         \label{Fig3}
   \end{figure*}

   For Blobs 1-3, we compared the results in their radial velocity maps (see Fig. \ref{Fig2}a and \ref{Fig2}c) with other methods. We calculated the mean value and standard deviation of the radial velocity inside the region of the blob (see the region enclosed by the dashed purple curves in Figure \ref{Fig2}a or \ref{Fig2}c) and display them in the top row of Figure \ref{Fig3} as crosses and error bars that agree well with each other for STEREO-A/B. The black squares in the top row show the radial velocity of the icecream ball center for the icecream cone model, which is derived by five-point running fitting to the height-time profiles. The velocities from the icecream cone model fluctuate strongly over a range of about 200 km/s because it is difficult to accurately identify the leading edge and rear part of the blobs, especially when the blobs fade out, but they are overall consistent with the velocities obtained with our CORAR technique. We also estimated the radial velocity of the background solar wind at the same position and at same time with the ENLIL model \citep[][the thick solid green line in the top row of Figure \ref{Fig3}]{odstrvcil1999three, odstrcil2009numerical}. For the ENLIL simulation, we used the magnetograms obtained by the National Solar Observatory as input, and the inner boundary solar wind condition at $22 R_{\odot}$ was inferred using the Wang-Sheely-Arge empirical model \citep{1990ApJ...355..726W}. The radial, angular, and temporal grid size of the simulation is about $1.3 R_{\odot} \times 4^{\circ} \times 6 h,$ and we interpolated the simulation result to the reconstructed pp of Blobs 1-3. These small transients are thought to be the tracers of the background solar wind and should have the same velocity. The simulated velocity at the position of Blob 1 agrees well with our results, while the simulated velocity of for Blobs 2 and 3 is either about $100$ km/s higher or lower than ours. This deviation probably results from the preceding CME, as well as from the expansion and interaction of the two blobs, which caused Blob 2 to be slightly faster, but Blob 3 to be slower. These factors are not taken into account in this ENLIL simulation.

   \subsection{Analysis of the radial velocity maps of the CME on 3-4 April 2010}

   For the velocity maps of the CME on 3-4 April 2010, the $v_{rB}$ and $v_{rF}$ are more consistent in the inner region of the CME ejecta than at the leading edge (see rows 3 and 4 of Fig. \ref{Fig2}b). \citet{ying2019first} also found that the derived radial velocity has a smaller relative deviation from its actual value in the CME inner region by applying their cross-correlation method on the synthetic images of a simulated CME. The main reason is perhaps that for the inner region of a CME, the sampling box is fully filled with the CME features, while near the boundary, the CME only occupies a part of the sampling box, leading to a poorer representation of the local CME density variation and the large uncertainty of the derived radial velocity. The maps of the final velocity, the average of $v_{rB}$ and $v_{rF}$ from STEREO-A and B, show the very similar spatial distributions between the leading edge of the ejecta ($>600$ km/s) and the core ($\approx500$ km/s) (see row 5 of Fig. \ref{Fig2}b), suggesting the expansion of the CME. The latitudinal velocity difference is also revealed: from nearly $900$ km/s in the northern part ($-5^\circ<\lambda<5^\circ$) to about $700$ km/s in the southern part ($-35^\circ<\lambda<-20^\circ$) at the leading edge of the ejecta, suggesting a distortion of the CME.

\begin{figure*}[htbp]
    \begin{center}
      \includegraphics[width=\hsize]{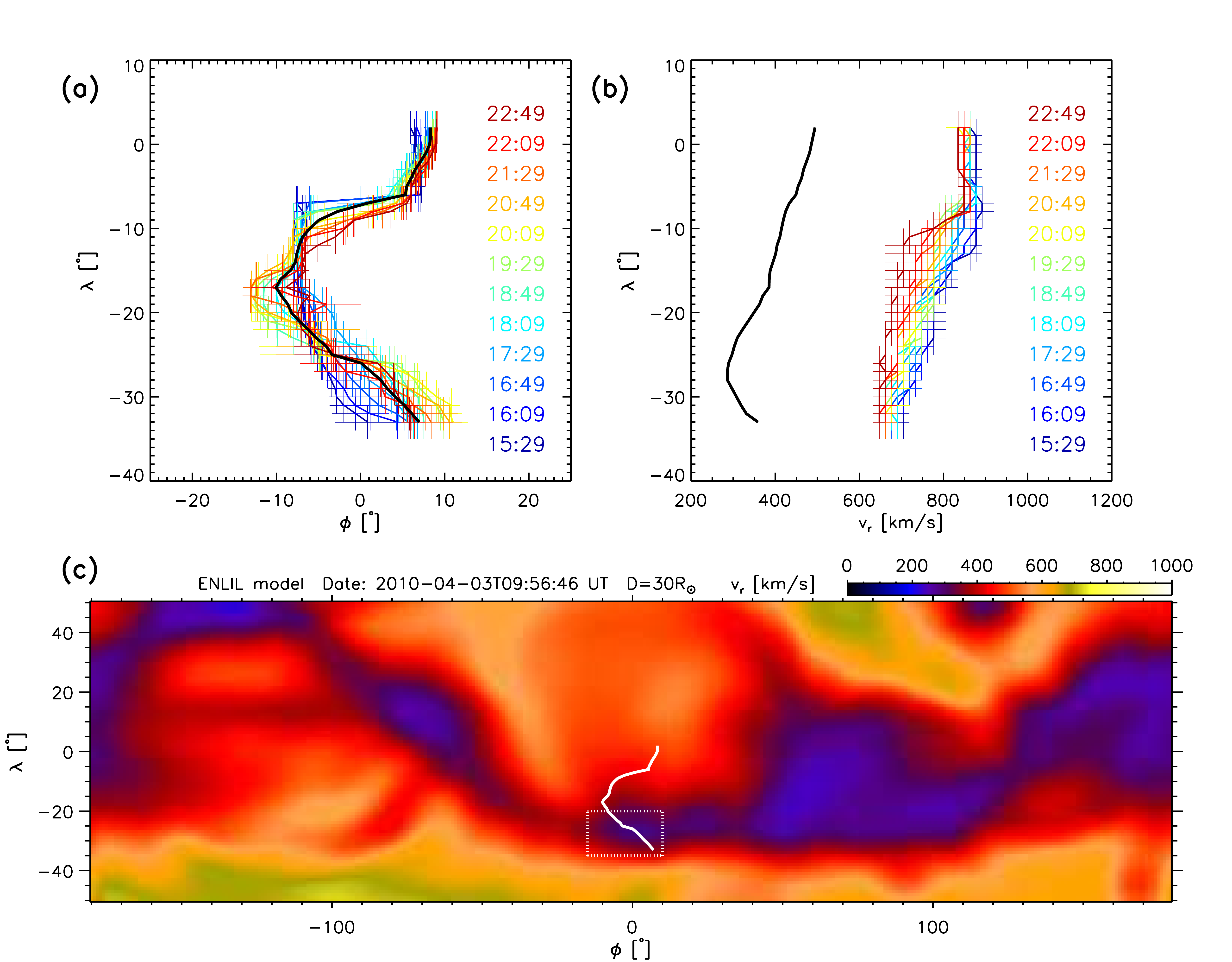}
    \end{center}
      \caption{Panels (a) and (b) show the angular position and the radial velocity of the ejecta leading edge derived from the 3D pp map and the radial velocity map, respectively, from 15:29 UT to 22:49 UT on 3 April 2010. Different colors represent different times. The thick black curve in Panel (a) is the temporally averaged HEE longitude ($\phi$) vs. HEE latitude ($\lambda$) of the CME front at this period of time. Panel (c) displays the angular distribution of the radial velocity of the simulated background solar wind in HEE coordinates with the ENLIL code at 9:56 UT on 3 April 2010 at a heliocentric distance of $30 R_{\odot}$. The dotted white box highlights the fast-to-slow streaming interaction region. The thick white curve corresponds to the thick black curve in Panel (a). The radial velocity of the background solar wind along the thick white line in Panel (c) is plotted in Panel (b) as the thick black curve.
              }
         \label{Fig4}
   \end{figure*}

   We further analyzed the propagation of the CME based on the derived leading-edge velocity. For the period between 15:29 UT to 22:49 UT on 3 April, when nearly the full ejecta leading edge could be reconstructed and accurately located with the CORAR method, we calculated the median values of the longitudes along the ejecta leading edge every $5^\circ$ in latitude for STEREO-A and B. Because the longitudes from STEREO-A are slightly different from those of STEREO-B, we treated them as the uncertainty and their mean values as the final longitudes, as shown by the color-coded lines in Figure \ref{Fig4}a. The thick black line denotes the temporal average of the leading edge of the ejecta. Repeating this for the radial velocity maps from STEREO-A/B, we obtain the variation of the radial velocity along the leading edge (see Fig. \ref{Fig4}b). Figure \ref{Fig4}b clearly shows that the CME was slowing down from $700 - 900$ km/s to $600 - 800$ km/s during the period. The derived ejecta leading-edge velocity stays within the measurement uncertainty of other methods near the ecliptic plane (see Table \ref{Tab1}). The radial velocity from catalog of the HEliospheric Cataloging, Analysis and Techniques Service (HELCATS) \citep{2013ApJ...777..167D} looks much higher than ours, perhaps because of the different propagating direction deductions of the HELCATS catalog, which seems too far east or too far west only in the HI images from STEREO-A or STEREO-B. The background solar wind speed map at $30 R_{\odot}$ at 9:56 UT on 3 April 2010 was calculated by using the ENLIL code and is shown in Figure \ref{Fig4}c . The curved leading edge of the ejecta is overplotted as the white line. Along the white line, the background solar wind ahead of the CME is obtained, which is shown in Figure \ref{Fig4}b as the thick black line. In Figure \ref{Fig4} the profile of the background solar wind speed matches the ejecta leading edge speed fairly well. The southern part of the CME has a lower speed because the speed of the ambient solar wind is lower. This result also supports results of \cite{2017SoPh..292..118S}, who suggested that the drag force from the background solar wind starts taking over the dynamics of this CME at a height of $5.5 R_{\odot}$.

\begin{figure*}[htbp]
    \begin{center}
      \includegraphics[width=\hsize]{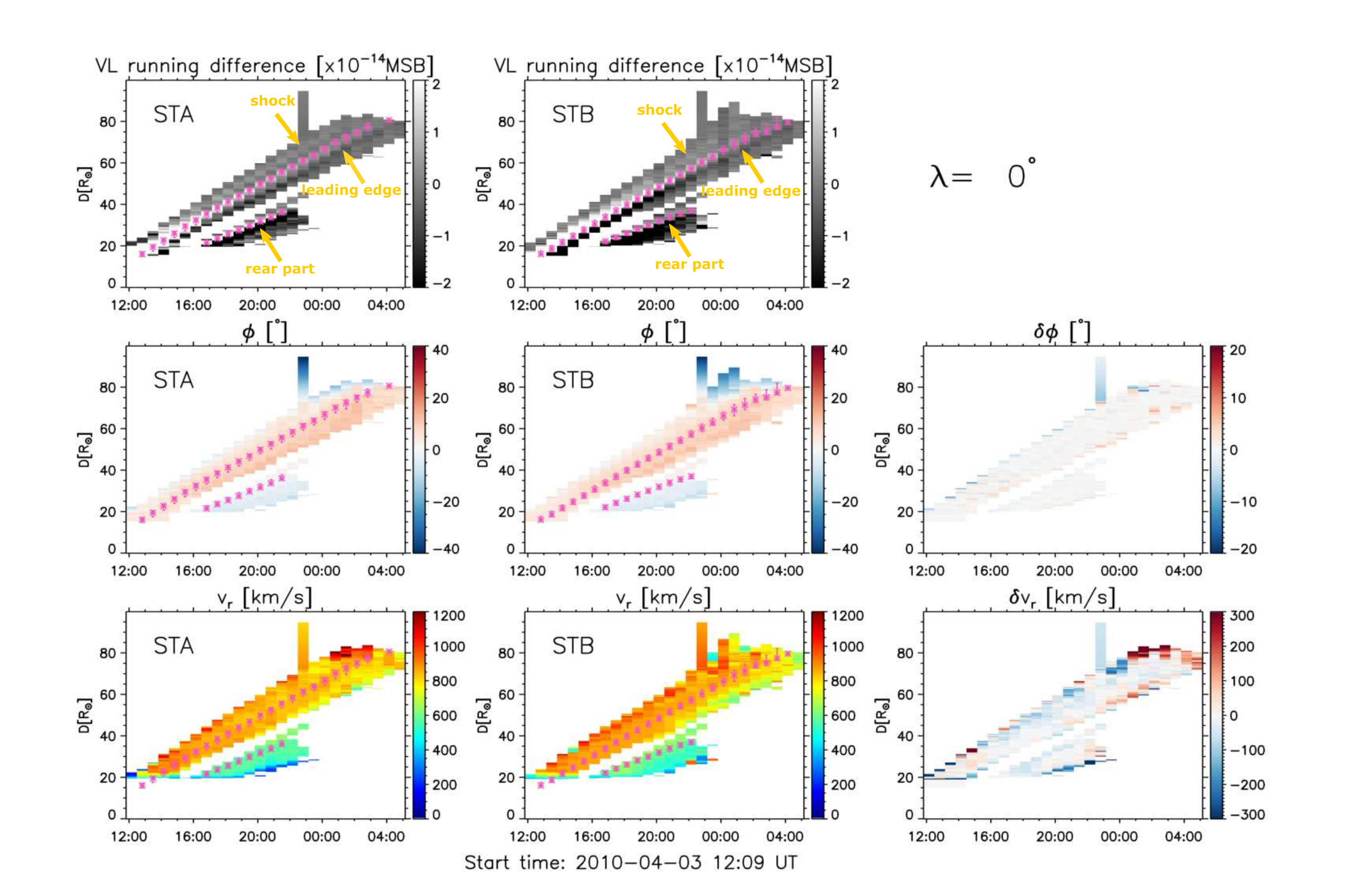}
    \end{center}
      \caption{Time-distance evolution of the CME on 3 April 2010 in the ecliptic plane ($\lambda=0^{\circ}$) derived from the 3D pp maps and radial velocity maps. The top, middle, and bottom rows display the time-distance plot of the brightness of STEREO/HI1 running-difference image, HEE longitude ($\phi$), and radial velocity ($v_r$), respectively. The left, middle, and right column represent the corresponding values of STEREO-A, STEREO-B, and their difference (A-B), respectively. The lilac asterisks with error bars mark the position of the ejecta leading part and rear part, identified by the peak of bright edges in the front and the back of the ejecta.
              }
         \label{Fig5}
   \end{figure*}

\begin{figure*}[htbp]
    \begin{center}
      \includegraphics[width=0.75\hsize]{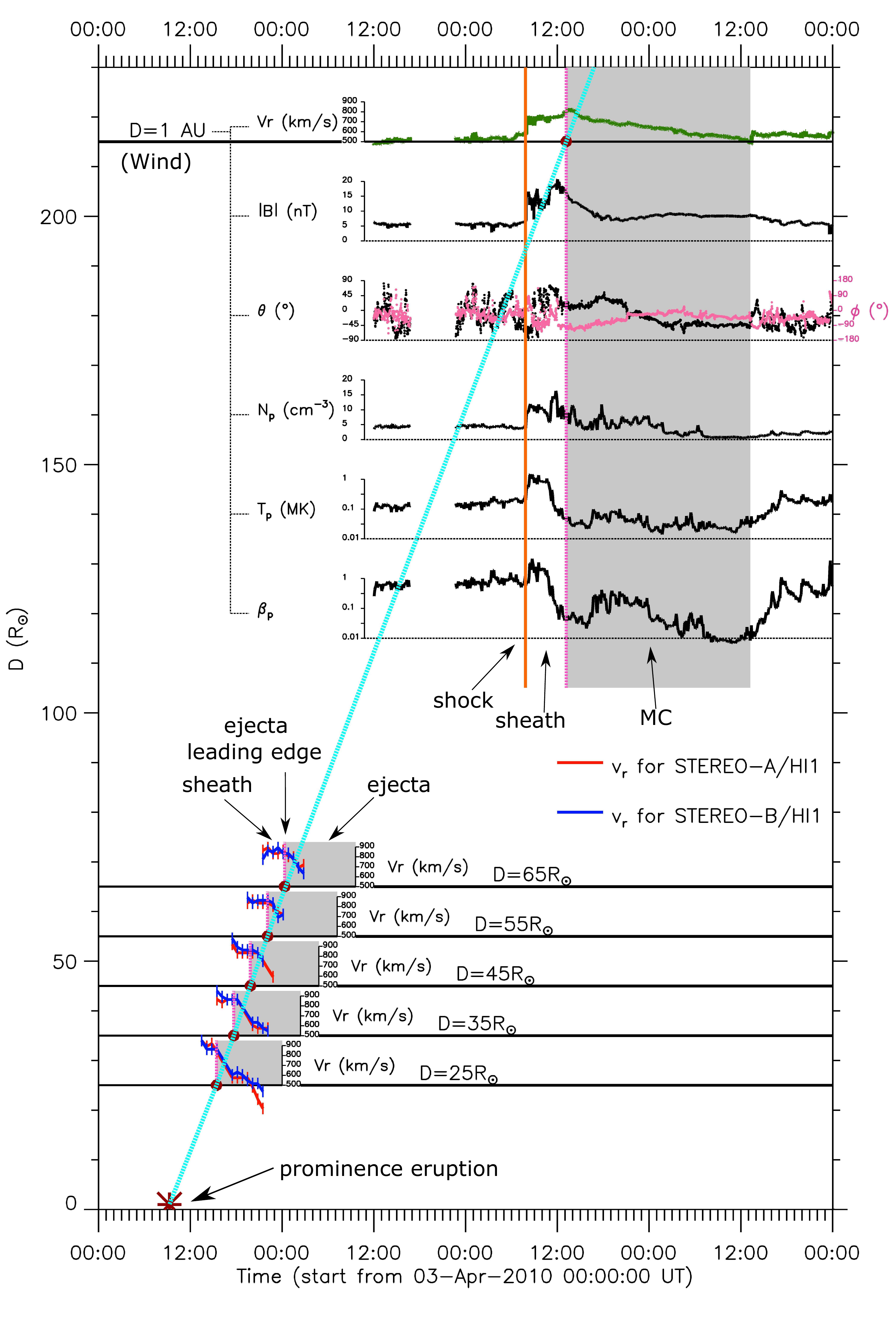}
    \end{center}
      \caption{Temporal evolution of the radial velocity of the CME with the different heliocentric distances along the Sun-Earth line. The horizontal axes of all the small insets are the time marked by the main horizontal axis at the bottom. The red (blue) thick lines with error bars represent the reconstructed solar wind radial velocity at the distance of $25 R_{\odot}$, $35 R_{\odot}$, $45 R_{\odot}$, $55 R_{\odot}$ , and $65 R_{\odot}$ for STEREO-A (B) based on the radial velocity map. The ejecta following its leading edge (vertical dashed lilac line) is marked as the shadow region. The brown asterisk above the main horizontal axis at the bottom represents the time of the prominence eruption associated with this CME. In the upper corner, the thick solid green line is the radial solar wind velocity observed by the Wind spacecraft at 1 AU. The vertical orange line and vertical dashed lilac line mark the arrivals of the shock and MC (in shadow), respectively (based on the list of Interplanetary Coronal Mass Ejections at http://space.ustc.edu.cn/dreams/wind$\_$icmes/index.php, \citeauthor{2016SoPh..291.2419C}, \citeyear{2016SoPh..291.2419C}). The magnetic field magnitude ($|B|$), elevation ($\theta$), and azimuth ($\phi$, in pink), the proton number density ($N_p$), the proton temperature ($T_p$), and the plasma beta ($\beta_p$) measured by Wind are also displayed for clarity. The oblique dashed blue polyline connects the leading edge of the ejecta in the HI1 FOV (solid brown circles at distances of $25 R_{\odot}$, $35 R_{\odot}$, $45 R_{\odot}$, $55 R_{\odot}$ , and $65 R_{\odot}$) with the CME eruption time at the solar surface (brown asterisk) and the beginning time of magnetic cloud at 1 AU (solid brown circle at 1 AU).
              }
         \label{Fig6}
   \end{figure*}

   To trace the velocity evolution inside the CME ejecta, we combined our MCT results on the ecliptic plane at a heliocentric distance of $20-60 R_{\odot}$ and the in situ observations at 1 AU. This CME was observed in situ by the Wind spacecraft near the Earth at 1 AU on 5 April 2010, which clearly show the signatures of a magnetic cloud (MC) \citep{2010GeoRL..3724103M}(see Fig. \ref{Fig6}). Based on the pp maps and velocity maps, we particularly analyzed the velocity evolution with the time and distance inside the CME (see the first row of Fig. \ref{Fig5}) on the ecliptic plane ($\lambda=0^\circ$). The probable longitude of the CME on the ecliptic plane is within $10^\circ$ (see row 2 of Fig. \ref{Fig5}), close to the Sun-Earth line. As shown in row 1 of Figure \ref{Fig5}, from the time-distance distribution of the running-difference HI1 images brightness, we can identify two features. One is the CME leading part, which contains the ejecta leading edge and the driven shock, and the other is the CME rear part. The velocity of the leading part remained at about $850$ km/s, while at the rear part of the ejecta, it decreased to about $500$ km/s. To compare with the in situ observations at 1 AU, the velocity profiles at distances of $25 R_{\odot}$, $35 R_{\odot}$, $45 R_{\odot}$, $55 R_{\odot}$ , and $65 R_{\odot}$ were extracted from the bottom row of Figure \ref{Fig5} and are shown in Figure \ref{Fig6}. The oblique dashed blue polyline in Figure \ref{Fig6} connects the leading edge of the ejecta in the HI1 FOV with the CME eruption time at the solar surface (i.e., the onset of the corresponding prominence eruption, which is introduced in detail in \citeauthor{2012JGRA..117.4105X}, \citeyear{2012JGRA..117.4105X}) and the beginning time of the MC at 1 AU (based on the list of Interplanetary Coronal Mass Ejections at http://space.ustc.edu.cn/dreams/wind$\_$icmes/index.php, \citeauthor{2016SoPh..291.2419C}, \citeyear{2016SoPh..291.2419C}). It is almost a straight line in the time-distance map, corresponding to a radial velocity of about 800 km/s, suggesting that the radial velocity map of the CME is well reconstructed with our method. The rear part became diffused and quickly unrecognizable in the HI1 FOV. We therefore did not repeat this procedure for the CME rear part.

\section{Application: Deriving the dimensionless expansion rate of a CME ejecta in white-light images}

   Coronal mass ejections may expand due to force balance between the pressure in the CME ejecta and the pressure of the ambient solar wind. With the solar wind pressure dropping following ${P_t \propto D^{-n_p}}$, where $D$ is distance from the Sun and $n_p$ is the decay index, \citet{demoulin2009causes} demonstrated theoretically that the MC flux rope radius evolves as $D^{n_p/4}$. For solar wind with $n_p$ of about $2.91\pm0.31$, the dimensionless expansion rate $\zeta=n_p/4=0.73\pm0.08$ \citep{gulisano2010global}.

   In reality, this expansion rate of CME ejecta can normally be estimated with in situ velocity observations. The estimated $\zeta$ varies from $0.95\pm1.05$ \citep{2020ApJ...899..119L}, $0.92\pm0.07$ \citep{2005P&SS...53....3L}, $0.80\pm0.20$ \citep{demoulin2008expected}, $0.78\pm0.10$ \citep{1998AnGeo..16....1B}, $0.6$ \citep{wang2005characteristics,leitner2007consequences,2015JGRA..120.1543W} to $0.38\pm1.08$ \citep{2019ApJ...877...77V} beyond $0.3$ AU. \citet{gulisano2010global} found that it was $0.91\pm0.23$ for unperturbed MCs and $0.48\pm0.79$ for perturbed MCs at a heliocentric distance between $0.3$ to $1$ AU. Because the value of $\zeta$ is generally between $0$ and $1$ and the CME normally maintains its angular size, CMEs usually become flatter or pancake during propagation \citep{1996AIPC..382..442C, 2002P&SS...50..527R, 2003JGRA..108.1272R, 2004JGRA..109.2107M, 2004ApJ...600.1035R}. \cite{2020ApJ...899..119L} pointed out that the in situ measurements at a certain distance do not reflect the expansion of CMEs at a closer heliocentric distance. As for heliocentric distance within $0.3$ AU (about $60 R_{\odot}$), currently only the Parker Solar Probe has the opportunity to collect in situ observation data \citep{2016SSRv..204..131K} to determine the expansion rate $\zeta$.

   With the 3D position and radial velocity reconstruction of a CME from the 2D white-light images, we can derive the expansion rate $\zeta$ at distance smaller than $60 R_{\odot}$. In addition, with the time cadence of $40$ minutes for STEREO/HI1 images, we can trace the variation of $\zeta$ in different parts of the same CME as it propagates out.

   \subsection{Method}

     \citet{wood2016imaging} developed a method for quantifying the self-similarity of the expansion of prominence structures in white-light images based on the reconstructed 3D position and velocity of the structure point to point in each frame. The self-similarity of the radial expansion is defined as
     \begin{eqnarray}
       S_r \equiv \frac{D}{a} \frac{da}{dD} = \frac{(D_1+D_2)(V_1-V_2)}{(D_1-D_2)(V_1+V_2)} \label{Sr definition}
     ,\end{eqnarray}
     where $a$ represents the half radial extent of the object, $D$ is the distance of the object center from the Sun, $D_1=D+a$, $D_2=D-a$ are the distance of the front and back of the object, and $V_1$ and $V_2$ are the radial velocity of the front and back of the object. For self-similar expansion, $S_r=1$. If the object expands faster or slower than self-similar expansions, $S_r>1$ or $<1$. If the object maintains its radial size or contracts, $S_r\le0$. Because $a \propto D^{\zeta}$ according to the definition of the dimensionless expansion rate $\zeta$, it is easy to deduce that
     \begin{eqnarray}
       S_r=\zeta
     ,\end{eqnarray}
     which means that $S_r$ and $\zeta$ are equivalent. We therefore applied their method to calculate the $S_r$ (i.e., $\zeta$) of different parts of the CME on 3 April 2010 with the 3D pp map and radial velocity map derived above.

     As \citet{wood2016imaging} discussed, for a structure to maintain the same shape relative to the Sun (i.e., $\zeta = 1$), the velocities within the structure in the direction of propagation must be proportional to distance from the Sun, but for a structure with $\zeta \ne 1$, the radial velocity would be proportional to the power law of the heliocentric distance. We therefore used the power-law fitting function
     \begin{eqnarray}
       V_r=k_1D^{k_2}
     \end{eqnarray}
      to fit the data points within certain parts of the ejecta in each frame. To reduce the fluctuation of the velocities of the data points over distance, we calculated the piecewise mean $v_r$ and piecewise standard deviation within every radial length of $0.5 R_{\odot}$ as the data points for fitting. After the fitting parameters $k_1$ and $k_2$ were obtained, we derived $V_1=k_1{D_1}^{k_2}$ and $V_2=k_1{D_2}^{k_2}$, and therefore $S_r$ (i.e., $\zeta$).

   \subsection{Results and analysis}

\begin{figure*}[htbp]
    \begin{center}

     \includegraphics[width=0.4\hsize]{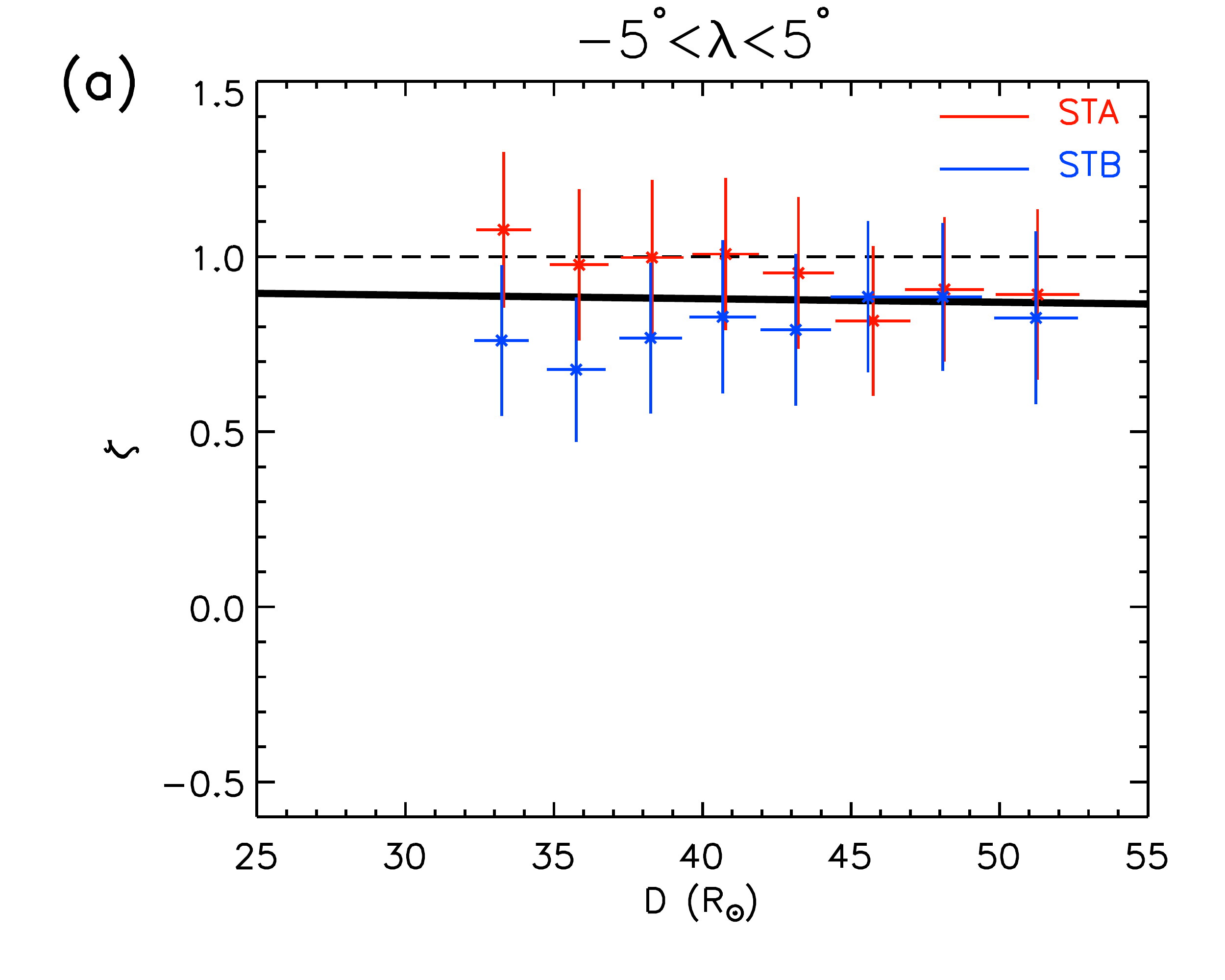}

     \includegraphics[width=0.4\hsize]{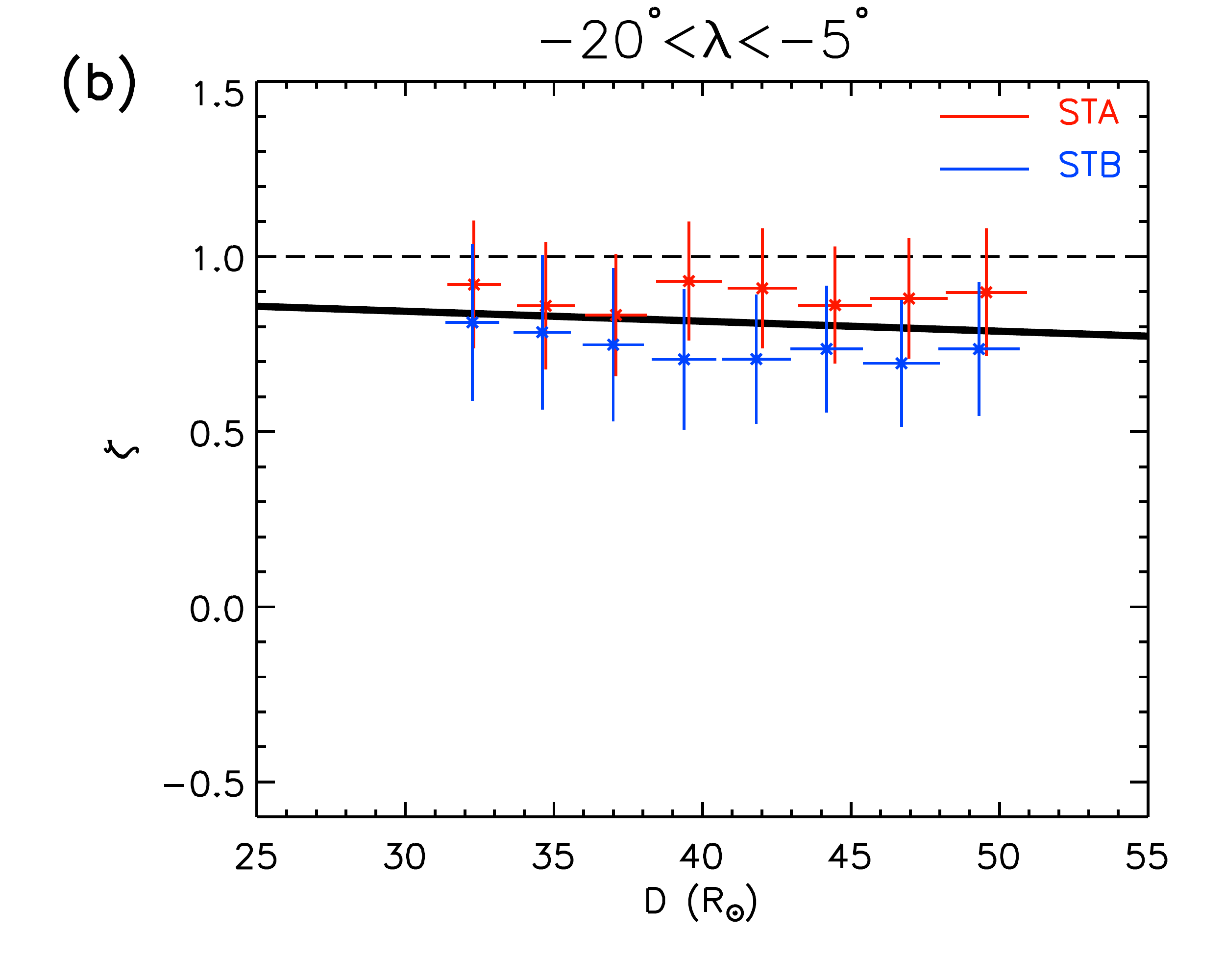}

     \includegraphics[width=0.4\hsize]{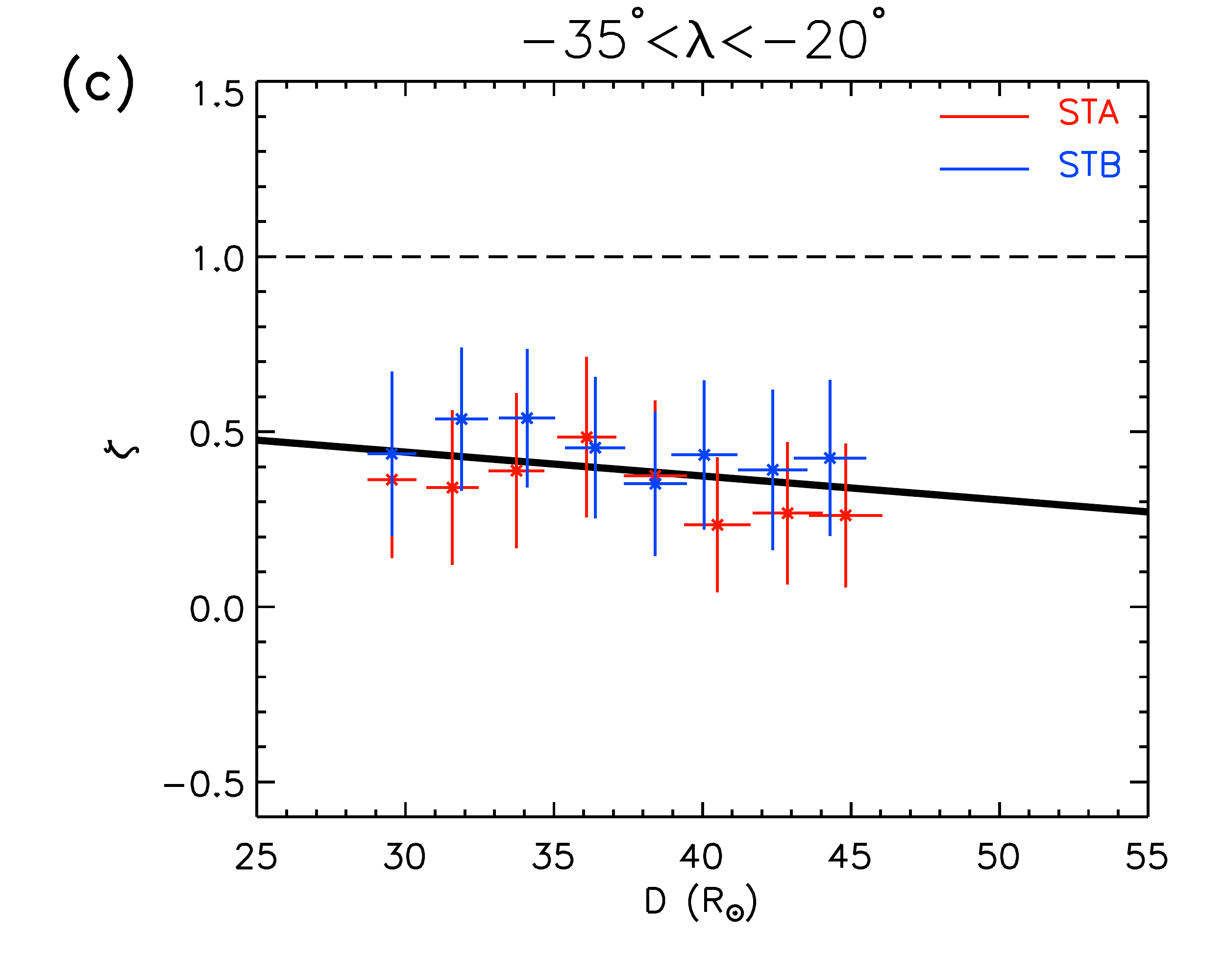}

    \end{center}
    \caption{Panels (a)-(c) show the dimensionless expansion rate ($\zeta$) vs. heliocentric distance ($D$) for the northern part ($-5^{\circ}<\lambda<5^{\circ}$), middle part ($-20^{\circ}<\lambda<-5^{\circ}$), and southern part ($-35^{\circ}<\lambda<-20^{\circ}$) of the CME ejecta, respectively, between 18:09 UT and 22:49 UT on 3 April 2010, where $\lambda$ is the latitude in the HEE coordinates. The red and blue crosses with the horizontal and vertical error bars correspond to STEREO-A/B, while the thick solid black line is the linear fitting of the data points. The dashed line shows the value of $\zeta$ equal to $1$.
      }
    \label{Fig7}
   \end{figure*}

     Figure \ref{Fig7} display the derived $\zeta$ versus $D$ plots for the northern part ($-5^\circ<\lambda<5^\circ$), middle part ($-20^\circ<\lambda<-5^\circ$), and southern part ($-35^\circ<\lambda<-20^\circ$) of the CME ejecta. The solid black line shows the linear fitting line. The values of $\zeta$ for STEREO-A (red) and STEREO-B (blue) are different, but the deviations are smaller than the uncertainty, and therefore they are consistent enough to reveal the expansion rate of the ejecta.

    For the northern and middle part of the CME, $\zeta$ keeps a high value of $0.87 \pm 0.11$ and $0.81 \pm 0.09,$  respectively, at a distance between $25R_{\odot}$ and $55R_{\odot}$ (see Fig. \ref{Fig7}a and \ref{Fig7}b). This suggests that the northern and middle parts of this CME maintained a nearly self-similar expansion but were still slowly pancaked.

    For the southern part of the CME, $\zeta$ is in the range of $0.39 \pm 0.09$ (see Fig. \ref{Fig7}c), which means a higher degree of flattening than in other parts. This is because the ambient solar wind was much slower than the CME, as shown in Figure \ref{Fig5}b, which acted as an obstacle and prevented the CME from expanding. A similarly low expansion rate ($0.48\pm0.79$) was reported by \citet{gulisano2010global}, but the difference is that the low expansion rate in their work is due to the compression by the faster background solar wind from the back.

   The question now is why the CME different latitudinal parts show different dimensionless expansion rates, or in broad terms, different dynamics. The CME on 3 April 2010 is not the only one with such diverse properties. \citet{savani2010observational} have previously found a similar differentiated radial distortion event on 2007 November 14 in which an initially circular CME was observed to be distorted into an increasingly concave structure in STEREO-HI1 images due to the latitudinally varied solar wind velocity ahead. In many magnetohydrodynamic simulations, the fast and slow solar wind ahead of a fast CME could also lead to a similar highly distorted concave shape of the CME at a heliocentric distance larger than $20R_{\odot}$ \citep[e.g., ][]{2004JGRA..109.2107M, 2005ApJ...627.1019L}. \citet{owens2017coronal} further pointed out that CMEs with constant angular width cease to be coherent magnetohydrodynamic structures within $0.3$ AU from the Sun because beyond a certain heliocentric distance, the information spreading speed (i.e., the Alfv{\'e}n wave speed $V_A$) in magnetized plasma decreases less than the geometric separation speed of two points on the ejecta leading edge. \citet{owens2017coronal} suggested that different parts of a CME may be isolated from each other and the CME interacting with the background solar wind may be similar to dust cloud rather than a billiard ball as a whole. In their work, $B$, the magnetic field intensity within the CME, $A$, the CME cross-sectional area, and $n$, the ion density at heliocentric distance $D$ are related like $B=B_0  \frac{A_0}{A}$ and $n=n_0  \frac{A_0 D_0}{AD}$ , where $B_0$, $A_0$ , and $n_0$ are the value of $B$, $A$ , and $n$ at $D_0$ of $1$ AU. Assuming the CME self-similar expands in angular space, then $A \propto aD \propto D^{\zeta+1}$ according to what we discussed in Section 5, where $a$ is half of the radial extent of the CME, while $\zeta$ is the dimensionless expansion rate. Inserting all the above into the formula of the Alfv{\'e}n wave speed $V_A=\frac{B}{\sqrt{\mu_0 n m_i}}$ , in which ${\mu_0}$ is the magnetic permeability of free space and $m_i$ is the mean ion mass, we deduce that $V_A=V_{A_0} {\frac{D_0}{D}}^{\frac{\zeta}{2}}$ where $V_{A_0}$ is the Alfv{\'e}n wave speed at $1$ AU. The geometric separation speed $V_G=V_{TR} \theta$ with $\theta$ as separation angle of two points and $V_{TR}$ as the radial velocity on the ejecta leading edge. When $V_G>V_A$, it can be derived that $D>D_0 {\frac{V_{A_0}}{V_{TR} \theta}}^{\frac{2}{\zeta}} $, so that the critical distance is
    \begin{eqnarray}
      D_T=D_0 {\frac{V_{A_0}}{V_{TR} \theta}}^{\frac{2}{\zeta}}.  \label{DT}
    \end{eqnarray}
    The CME on 3 April 2010 was observed by the Wind spacecraft near the Earth at 1 AU. According to the Wind in situ measurements of the interplanetary magnetic field and the solar wind plasma, at the arrival of the MC, $B_0$ is about $15$ nT, $n_0$ about $10$ $cm^{-3}$ (see Fig. \ref{Fig6}), and therefore $V_{A_0}$ is about $104$ km/s. According to our reconstruction of this CME, near the ecliptic plane, $V_{TR} \approx 800$ km/s, $\zeta \approx 0.9$. Assuming that the CME kept a constant angular width, based on Eq. \ref{DT}, we can derive that for the whole CME, $\theta \approx 40^{\circ}$ and $D_T\approx5 R_{\odot}$, while for the northern, middle, and southern parts, the separation angle between them is about $15^{\circ}$ and $D_T\approx43 R_{\odot}$, which indicates that in the HI1 FOV, this CME was already incoherent and its northern, middle, and southern parts evolved independently.

  \section{Summary and discussion}

    We developed an MCT method to reconstruct the radial velocity distribution inside solar wind transients in 3D space based on STEREO HI1 images. The MCT technique has two steps: maximum cc localization, and cross-correlation tracking. The velocity estimated with the MCT technique is the velocity that is normally temporally averaged over 40 minutes. The uncertainty is smaller than 50 km/s and is mainly caused by the step size of the radial shift, which is limited by the spatial resolution of the HI1 images. For slow transients, this uncertainty is notable. In addition, limited by the weakness of the CORAR method, such as the collinear effect (see Section 3.3 of Paper II for details), some solar wind transients may not be well or completely reconstructed, leading to the deviations or null regions in the 3D pp map and radial velocity map.

    Although the CORAR and MCT techniques have some weaknesses, by applying them on the HI1 image data on 3-4 April 2010, we find that the derived position and velocity information match the results from a forward-modeling method and the in situ observations at $1$ AU well. The main results on the events are listed below.

    1. The three blobs propagated outward with a constant radial velocity of about $400$, $500,$ and $400$ km/s along the direction around $S18^\circ W13^\circ$, $S19^\circ W11^\circ,$ and $S24^\circ W25^\circ$ in HEE coordinates, respectively.

    2. The radial velocity of the CME ejecta varied from over $600$ km/s at the leading edge to about $500$ km/s at the core, suggesting an expansion.

    3. The ambient solar wind distorted and shaped the ejecta front, causing its leading edge velocity in the northern part to be about 900 km/s and that in the southern part to be about 700 km/s.

    4. The dimensionless radial expansion rate of the CME is found to be $0.7 - 1.0$ in the northern and middle parts and $0.3-0.5$ in the southern part between a heliocentric distance of about $25 R_{\odot}$ and $55 R_{\odot}$, suggesting an incoherent evolution of the CME caused by the ambient solar wind.

    Small solar wind transients such as the blobs in white-light images are ideal tracers of the slow solar wind \citep{1997Measurements, 2017ApJ...851...32S, L2018Deprojected}. From the
    3D pp map and radial velocity maps of the three blob-like small transients, the velocity of the ambient slow solar wind can be detected. Because small transients are more frequently observed in coronagraphs or heliospheric images than large transients, it is possible to generate a velocity map of the global slow solar wind in the 3D inner heliosphere with the MCT method. With more observations from other perspectives, such as WISPR of PSP \citep{2016SSRv..204...83V}, SoloHI of Solar Orbiter \citep{howard2019solar, 2020A&A...642A...1M}, or the white-light imagers of the future possible Solar Ring mission \citep{2020ScChE..63.1699W}, the technique developed in this work can be used to better understand solar wind transients and the inner heliosphere.

\begin{acknowledgements}
     The STEREO/SECCHI data are produced by a consortium of NRL (USA), RAL (UK), LMSAL (USA), GSFC (USA), MPS (Germany), CSL (Belgium), IOTA (France), and IAS (France). The SOHO/LASCO data are produced by a consortium of the Naval Research Laboratory (USA), Max-Planck-Institut f\"{u}r Aeronomie (Germany), Laboratoire d'Astronomie (France), and the University of Birmingham (UK). The SECCHI data presented in this paper were obtained from STEREO Science Center (https://stereo-ssc.nascom.nasa.gov/data/ins\_data/secchi\_hi/L2). The Wind data were obtained from the Space Physics Data Facility (https://cdaweb.sci.gsfc.nasa.gov/). Simulation results have been provided by the Community Coordinated Modeling Center at Goddard Space Flight Center through their public Runs on Request system (http://ccmc.gsfc.nasa.gov). The ENLIL model was developed by D. Odstrcil at the University of Colorado at Boulder. We acknowledge the use of them. We acknowledge for the storage support from National Space Science Data Center, National Science and Technology Infrastructure of China (http://www.nssdc.ac.cn). This work is supported by the grants from the Strategic Priority Program of the Chinese Academy of Sciences (XDB41000000 and XDA15017300), the NSFC (41774178 and 41761134088) and the fundamental research funds for the central universities (WK2080000077). Y.W. is particularly grateful to the support of the Tencent Foundation.
\end{acknowledgements}

  \bibliographystyle{aa}
  \bibliography{Paper4}

\begin{thebibliography}{80}
\expandafter\ifx\csname natexlab\endcsname\relax\def\natexlab#1{#1}\fi

\bibitem[{{Barnes} {et~al.}(2019){Barnes}, {Davies}, {Harrison}, {Byrne},
  {Perry}, {Bothmer}, {Eastwood}, {Gallagher}, {Kilpua}, {M{\"o}stl},
  {Rodriguez}, {Rouillard}, \& {Odstr{\v{c}}il}}]{2019SoPh..294...57B}
{Barnes}, D., {Davies}, J.~A., {Harrison}, R.~A., {et~al.} 2019, \solphys, 294,
  57

\bibitem[{Bemporad {et~al.}(2018)Bemporad, Pagano, \&
  Giordano}]{bemporad2018measuring}
Bemporad, A., Pagano, P., \& Giordano, S. 2018, Astronomy \& Astrophysics, 619,
  A25

\bibitem[{{Bothmer} \& {Schwenn}(1998)}]{1998AnGeo..16....1B}
{Bothmer}, V. \& {Schwenn}, R. 1998, Annales Geophysicae, 16, 1

\bibitem[{Brueckner {et~al.}(1995)Brueckner, Howard, Koomen, Korendyke, \&
  Eyles}]{1995The}
Brueckner, G.~E., Howard, R.~A., Koomen, M.~J., Korendyke, C.~M., \& Eyles,
  C.~J. 1995, Solar Physics, 162, 357

\bibitem[{{Chen} {et~al.}(1997){Chen}, {Howard}, {Brueckner}, {Santoro},
  {Krall}, {Paswaters}, {St. Cyr}, {Schwenn}, {Lamy}, \&
  {Simnett}}]{1997ApJ...490L.191C}
{Chen}, J., {Howard}, R.~A., {Brueckner}, G.~E., {et~al.} 1997, \apjl, 490,
  L191

\bibitem[{{Chi} {et~al.}(2016){Chi}, {Shen}, {Wang}, {Xu}, {Ye}, \&
  {Wang}}]{2016SoPh..291.2419C}
{Chi}, Y., {Shen}, C., {Wang}, Y., {et~al.} 2016, \solphys, 291, 2419

\bibitem[{Colaninno \& Vourlidas(2006)}]{colaninno2006analysis}
Colaninno, R.~C. \& Vourlidas, A. 2006, The Astrophysical Journal, 652, 1747

\bibitem[{{Cremades} \& {Bothmer}(2004)}]{2004A&A...422..307C}
{Cremades}, H. \& {Bothmer}, V. 2004, \aap, 422, 307

\bibitem[{{Crooker} \& {Intriligator}(1996)}]{1996AIPC..382..442C}
{Crooker}, N.~U. \& {Intriligator}, D.~S. 1996, in American Institute of
  Physics Conference Series, Vol. 382, Proceedings of the eigth International
  solar wind Conference: Solar wind eight, ed. D.~{Winterhalter}, J.~T.
  {Gosling}, S.~R. {Habbal}, W.~S. {Kurth}, \& M.~{Neugebauer}, 442--444

\bibitem[{{Davies} {et~al.}(2012){Davies}, {Harrison}, {Perry}, {M{\"o}stl},
  {Lugaz}, {Rollett}, {Davis}, {Crothers}, {Temmer}, {Eyles}, \&
  {Savani}}]{2012ApJ...750...23D}
{Davies}, J.~A., {Harrison}, R.~A., {Perry}, C.~H., {et~al.} 2012, \apj, 750,
  23

\bibitem[{{Davies} {et~al.}(2013){Davies}, {Perry}, {Trines}, {Harrison},
  {Lugaz}, {M{\"o}stl}, {Liu}, \& {Steed}}]{2013ApJ...777..167D}
{Davies}, J.~A., {Perry}, C.~H., {Trines}, R.~M.~G.~M., {et~al.} 2013, \apj,
  777, 167

\bibitem[{DeForest {et~al.}(2017)DeForest, de~Koning, \&
  Elliott}]{deforest20173d}
DeForest, C., de~Koning, C., \& Elliott, H. 2017, The Astrophysical Journal,
  850, 130

\bibitem[{{DeForest} {et~al.}(2018){DeForest}, {Howard}, {Velli}, {Viall}, \&
  {Vourlidas}}]{2018ApJ...862...18D}
{DeForest}, C.~E., {Howard}, R.~A., {Velli}, M., {Viall}, N., \& {Vourlidas},
  A. 2018, \apj, 862, 18

\bibitem[{D{\'e}moulin \& Dasso(2009)}]{demoulin2009causes}
D{\'e}moulin, P. \& Dasso, S. 2009, Astronomy \& Astrophysics, 498, 551

\bibitem[{D{\'e}moulin {et~al.}(2008)D{\'e}moulin, Nakwacki, Dasso, \&
  Mandrini}]{demoulin2008expected}
D{\'e}moulin, P., Nakwacki, M.~S., Dasso, S., \& Mandrini, C.~H. 2008, Solar
  Physics, 250, 347

\bibitem[{Eyles {et~al.}(2009)Eyles, Harrison, Davis, Waltham, Shaughnessy,
  Mapson-Menard, Bewsher, Crothers, Davies, Simnett,
  {et~al.}}]{eyles2009heliospheric}
Eyles, C., Harrison, R., Davis, C.~J., {et~al.} 2009, Solar Physics, 254, 387

\bibitem[{Feng {et~al.}(2015)Feng, Inhester, \& Gan}]{feng2015radial}
Feng, L., Inhester, B., \& Gan, W. 2015, The Astrophysical Journal, 805, 113

\bibitem[{Feng {et~al.}(2012)Feng, Inhester, Wei, Gan, Zhang, \&
  Wang}]{feng2012morphological}
Feng, L., Inhester, B., Wei, Y., {et~al.} 2012, The Astrophysical Journal, 751,
  18

\bibitem[{{Fisher} \& {Munro}(1984)}]{1984ApJ...280..428F}
{Fisher}, R.~R. \& {Munro}, R.~H. 1984, \apj, 280, 428

\bibitem[{Gulisano {et~al.}(2010)Gulisano, D{\'e}moulin, Dasso, Ruiz, \&
  Marsch}]{gulisano2010global}
Gulisano, A.~M., D{\'e}moulin, P., Dasso, S., Ruiz, M.~E., \& Marsch, E. 2010,
  Astronomy \& Astrophysics, 509, A39

\bibitem[{Harrison {et~al.}(2008)Harrison, Davis, Eyles, Bewsher, Crothers,
  Davies, Howard, Moses, Socker, Newmark, {et~al.}}]{harrison2008first}
Harrison, R.~A., Davis, C.~J., Eyles, C.~J., {et~al.} 2008, Solar Physics, 247,
  171

\bibitem[{{Howard} {et~al.}(2008){Howard}, {Moses}, {Vourlidas}, {Newmark},
  {Socker}, {Plunkett}, {Korendyke}, {Cook}, {Hurley}, {Davila}, {Thompson},
  {St Cyr}, {Mentzell}, {Mehalick}, {Lemen}, {Wuelser}, {Duncan}, {Tarbell},
  {Wolfson}, {Moore}, {Harrison}, {Waltham}, {Lang}, {Davis}, {Eyles},
  {Mapson-Menard}, {Simnett}, {Halain}, {Defise}, {Mazy}, {Rochus}, {Mercier},
  {Ravet}, {Delmotte}, {Auchere}, {Delaboudiniere}, {Bothmer}, {Deutsch},
  {Wang}, {Rich}, {Cooper}, {Stephens}, {Maahs}, {Baugh}, {McMullin}, \&
  {Carter}}]{2008SSRv..136...67H}
{Howard}, R.~A., {Moses}, J.~D., {Vourlidas}, A., {et~al.} 2008, \ssr, 136, 67

\bibitem[{Howard \& Rochus(2019)}]{howard2019solar}
Howard, R.~A. \& Rochus, P. 2019, Astronomy and Astrophysics

\bibitem[{{Howard} \& {DeForest}(2012)}]{2012ApJ...752..130H}
{Howard}, T.~A. \& {DeForest}, C.~E. 2012, \apj, 752, 130

\bibitem[{{Howard} {et~al.}(2006){Howard}, {Webb}, {Tappin}, {Mizuno}, \&
  {Johnston}}]{2006JGRA..111.4105H}
{Howard}, T.~A., {Webb}, D.~F., {Tappin}, S.~J., {Mizuno}, D.~R., \&
  {Johnston}, J.~C. 2006, Journal of Geophysical Research (Space Physics), 111,
  A04105

\bibitem[{Jackson {et~al.}(1998)Jackson, Hick, Kojima, \&
  Yokobe}]{Jackson1998Heliospheric}
Jackson, B.~V., Hick, P.~L., Kojima, M., \& Yokobe, A. 1998, Journal of
  Geophysical Research Space Physics, 103, 12049

\bibitem[{{Kasper} {et~al.}(2016){Kasper}, {Abiad}, {Austin}, {Balat-Pichelin},
  {Bale}, {Belcher}, {Berg}, {Bergner}, {Berthomier}, {Bookbinder}, {Brodu},
  {Caldwell}, {Case}, {Chandran}, {Cheimets}, {Cirtain}, {Cranmer}, {Curtis},
  {Daigneau}, {Dalton}, {Dasgupta}, {DeTomaso}, {Diaz-Aguado}, {Djordjevic},
  {Donaskowski}, {Effinger}, {Florinski}, {Fox}, {Freeman}, {Gallagher},
  {Gary}, {Gauron}, {Gates}, {Goldstein}, {Golub}, {Gordon}, {Gurnee}, {Guth},
  {Halekas}, {Hatch}, {Heerikuisen}, {Ho}, {Hu}, {Johnson}, {Jordan},
  {Korreck}, {Larson}, {Lazarus}, {Li}, {Livi}, {Ludlam}, {Maksimovic},
  {McFadden}, {Marchant}, {Maruca}, {McComas}, {Messina}, {Mercer}, {Park},
  {Peddie}, {Pogorelov}, {Reinhart}, {Richardson}, {Robinson}, {Rosen},
  {Skoug}, {Slagle}, {Steinberg}, {Stevens}, {Szabo}, {Taylor}, {Tiu}, {Turin},
  {Velli}, {Webb}, {Whittlesey}, {Wright}, {Wu}, \&
  {Zank}}]{2016SSRv..204..131K}
{Kasper}, J.~C., {Abiad}, R., {Austin}, G., {et~al.} 2016, \ssr, 204, 131

\bibitem[{{Kilpua} {et~al.}(2009){Kilpua}, {Luhmann}, {Gosling}, {Li},
  {Elliott}, {Russell}, {Jian}, {Galvin}, {Larson}, {Schroeder}, {Simunac}, \&
  {Petrie}}]{2009SoPh..256..327K}
{Kilpua}, E.~K.~J., {Luhmann}, J.~G., {Gosling}, J., {et~al.} 2009, \solphys,
  256, 327

\bibitem[{{Krall} \& {St. Cyr}(2006)}]{2006ApJ...652.1740K}
{Krall}, J. \& {St. Cyr}, O.~C. 2006, \apj, 652, 1740

\bibitem[{Leitner {et~al.}(2007)Leitner, Farrugia, M{\"o}stl, Ogilvie, Galvin,
  Schwenn, \& Biernat}]{leitner2007consequences}
Leitner, M., Farrugia, C., M{\"o}stl, C., {et~al.} 2007, Journal of Geophysical
  Research: Space Physics, 112

\bibitem[{Li {et~al.}(2020)Li, Wang, Liu, Shen, Zhang, Lyu, Zhuang, Shen, Liu,
  \& Chi}]{2020Reconstructing}
Li, X., Wang, Y., Liu, R., {et~al.} 2020, Journal of Geophysical Research:
  Space Physics, 125

\bibitem[{Li {et~al.}(2018)Li, Wang, Liu, Shen, Zhang, Zhuang, Liu, \&
  Chi}]{li2018reconstructing}
Li, X., Wang, Y., Liu, R., {et~al.} 2018, Journal of Geophysical Research:
  Space Physics, 123, 7257

\bibitem[{Liewer {et~al.}(2011)Liewer, Hall, Howard, De~Jong, Thompson, \&
  Thernisien}]{liewer2011stereoscopic}
Liewer, P., Hall, J., Howard, R., {et~al.} 2011, Journal of atmospheric and
  solar-terrestrial physics, 73, 1173

\bibitem[{{Liu} {et~al.}(2010){Liu}, {Davies}, {Luhmann}, {Vourlidas}, {Bale},
  \& {Lin}}]{2010ApJ...710L..82L}
{Liu}, Y., {Davies}, J.~A., {Luhmann}, J.~G., {et~al.} 2010, \apjl, 710, L82

\bibitem[{{Liu} {et~al.}(2011){Liu}, {Luhmann}, {Bale}, \&
  {Lin}}]{2011ApJ...734...84L}
{Liu}, Y., {Luhmann}, J.~G., {Bale}, S.~D., \& {Lin}, R.~P. 2011, \apj, 734, 84

\bibitem[{{Liu} {et~al.}(2005){Liu}, {Richardson}, \&
  {Belcher}}]{2005P&SS...53....3L}
{Liu}, Y., {Richardson}, J.~D., \& {Belcher}, J.~W. 2005, \planss, 53, 3

\bibitem[{L{\'o}pez-Portela {et~al.}(2018)L{\'o}pez-Portela, Panasenco,
  Blanco-Cano, \& Stenborg}]{L2018Deprojected}
L{\'o}pez-Portela, C., Panasenco, O., Blanco-Cano, X., \& Stenborg, G. 2018,
  Solar Physics, 293, 99

\bibitem[{{Lugaz} {et~al.}(2005){Lugaz}, {Manchester}, \&
  {Gombosi}}]{2005ApJ...627.1019L}
{Lugaz}, N., {Manchester}, W.~B., I., \& {Gombosi}, T.~I. 2005, \apj, 627, 1019

\bibitem[{{Lugaz} {et~al.}(2020){Lugaz}, {Salman}, {Winslow}, {Al-Haddad},
  {Farrugia}, {Zhuang}, \& {Galvin}}]{2020ApJ...899..119L}
{Lugaz}, N., {Salman}, T.~M., {Winslow}, R.~M., {et~al.} 2020, \apj, 899, 119

\bibitem[{{Lyu} {et~al.}(2020){Lyu}, {Li}, \& {Wang}}]{2020AdSpR..66.2251L}
{Lyu}, S., {Li}, X., \& {Wang}, Y. 2020, Advances in Space Research, 66, 2251

\bibitem[{{Lyu} {et~al.}(2021){Lyu}, {Wang}, {Li}, {Guo}, {Wang}, \&
  {Zhang}}]{2021arXiv210103276L}
{Lyu}, S., {Wang}, Y., {Li}, X., {et~al.} 2021, arXiv e-prints,
  arXiv:2101.03276

\bibitem[{{Manchester} {et~al.}(2004){Manchester}, {Gombosi}, {Roussev},
  {Ridley}, {de Zeeuw}, {Sokolov}, {Powell}, \&
  {T{\'o}th}}]{2004JGRA..109.2107M}
{Manchester}, W.~B., {Gombosi}, T.~I., {Roussev}, I., {et~al.} 2004, Journal of
  Geophysical Research (Space Physics), 109, A02107

\bibitem[{Manoharan(2010)}]{2010Ooty}
Manoharan, P.~K. 2010, Solar Physics, 265, 137

\bibitem[{Mierla {et~al.}(2009)Mierla, Inhester, Marqu{\'e}, Rodriguez, Gissot,
  Zhukov, Berghmans, \& Davila}]{mierla20093d}
Mierla, M., Inhester, B., Marqu{\'e}, C., {et~al.} 2009, Solar Physics, 259,
  123

\bibitem[{Moran \& Davila(2004)}]{moran2004three}
Moran, T.~G. \& Davila, J.~M. 2004, Science, 305, 66

\bibitem[{{M{\"o}stl} {et~al.}(2014){M{\"o}stl}, {Amla}, {Hall}, {Liewer}, {De
  Jong}, {Colaninno}, {Veronig}, {Rollett}, {Temmer}, {Peinhart}, {Davies},
  {Lugaz}, {Liu}, {Farrugia}, {Luhmann}, {Vr{\v{s}}nak}, {Harrison}, \&
  {Galvin}}]{2014ApJ...787..119M}
{M{\"o}stl}, C., {Amla}, K., {Hall}, J.~R., {et~al.} 2014, \apj, 787, 119

\bibitem[{M{\"o}stl \& Davies(2013)}]{mostl2013speeds}
M{\"o}stl, C. \& Davies, J.~A. 2013, Solar Physics, 285, 411

\bibitem[{{M{\"o}stl} {et~al.}(2010){M{\"o}stl}, {Temmer}, {Rollett},
  {Farrugia}, {Liu}, {Veronig}, {Leitner}, {Galvin}, \&
  {Biernat}}]{2010GeoRL..3724103M}
{M{\"o}stl}, C., {Temmer}, M., {Rollett}, T., {et~al.} 2010, \grl, 37, L24103

\bibitem[{{M{\"u}ller} {et~al.}(2020){M{\"u}ller}, {St. Cyr}, {Zouganelis},
  {Gilbert}, {Marsden}, {Nieves-Chinchilla}, {Antonucci}, {Auch{\`e}re},
  {Berghmans}, {Horbury}, {Howard}, {Krucker}, {Maksimovic}, {Owen}, {Rochus},
  {Rodriguez-Pacheco}, {Romoli}, {Solanki}, {Bruno}, {Carlsson}, {Fludra},
  {Harra}, {Hassler}, {Livi}, {Louarn}, {Peter}, {Sch{"u}hle}, {Teriaca}, {del
  Toro Iniesta}, {Wimmer-Schweingruber}, {Marsch}, {Velli}, {De Groof},
  {Walsh}, \& {Williams}}]{2020A&A...642A...1M}
{M{\"u}ller}, D., {St. Cyr}, O.~C., {Zouganelis}, I., {et~al.} 2020, \aap, 642,
  A1

\bibitem[{Odstr{\v{c}}il \& Pizzo(1999)}]{odstrvcil1999three}
Odstr{\v{c}}il, D. \& Pizzo, V. 1999, Journal of Geophysical Research: Space
  Physics, 104, 483

\bibitem[{Odstrcil \& Pizzo(2009)}]{odstrcil2009numerical}
Odstrcil, D. \& Pizzo, V.~J. 2009, Solar Physics, 259, 297

\bibitem[{Ogilvie {et~al.}(1995)Ogilvie, Chornay, Fritzenreiter, Hunsaker,
  Keller, Lobell, Miller, Scudder, Sittler, \& Torbert}]{1995SWE}
Ogilvie, K.~W., Chornay, D.~J., Fritzenreiter, R.~J., {et~al.} 1995, Space .
  Rev, 71, 55

\bibitem[{Owens {et~al.}(2017)Owens, Lockwood, \& Barnard}]{owens2017coronal}
Owens, M., Lockwood, M., \& Barnard, L. 2017, Scientific Reports, 7, 1

\bibitem[{{Riley} \& {Crooker}(2004)}]{2004ApJ...600.1035R}
{Riley}, P. \& {Crooker}, N.~U. 2004, \apj, 600, 1035

\bibitem[{{Riley} {et~al.}(2003){Riley}, {Linker}, {Miki{\'c}}, {Odstrcil},
  {Zurbuchen}, {Lario}, \& {Lepping}}]{2003JGRA..108.1272R}
{Riley}, P., {Linker}, J.~A., {Miki{\'c}}, Z., {et~al.} 2003, Journal of
  Geophysical Research (Space Physics), 108, 1272

\bibitem[{{Rollett} {et~al.}(2012){Rollett}, {M{\"o}stl}, {Temmer}, {Veronig},
  {Farrugia}, \& {Biernat}}]{2012SoPh..276..293R}
{Rollett}, T., {M{\"o}stl}, C., {Temmer}, M., {et~al.} 2012, \solphys, 276, 293

\bibitem[{{Rouillard} {et~al.}(2011{\natexlab{a}}){Rouillard},
  {Odst{\v{r}}cil}, {Sheeley}, {Tylka}, {Vourlidas}, {Mason}, {Wu}, {Savani},
  {Wood}, {Ng}, {Stenborg}, {Szabo}, \& {St. Cyr}}]{2011ApJ...735....7R}
{Rouillard}, A.~P., {Odst{\v{r}}cil}, D., {Sheeley}, N.~R., {et~al.}
  2011{\natexlab{a}}, \apj, 735, 7

\bibitem[{{Rouillard} {et~al.}(2011{\natexlab{b}}){Rouillard}, {Sheeley},
  {Cooper}, {Davies}, {Lavraud}, {Kilpua}, {Skoug}, {Steinberg}, {Szabo},
  {Opitz}, \& {Sauvaud}}]{2011ApJ...734....7R}
{Rouillard}, A.~P., {Sheeley}, N.~R., J., {Cooper}, T.~J., {et~al.}
  2011{\natexlab{b}}, \apj, 734, 7

\bibitem[{{Russell} \& {Mulligan}(2002)}]{2002P&SS...50..527R}
{Russell}, C.~T. \& {Mulligan}, T. 2002, \planss, 50, 527

\bibitem[{{Sachdeva} {et~al.}(2017){Sachdeva}, {Subramanian}, {Vourlidas}, \&
  {Bothmer}}]{2017SoPh..292..118S}
{Sachdeva}, N., {Subramanian}, P., {Vourlidas}, A., \& {Bothmer}, V. 2017,
  \solphys, 292, 118

\bibitem[{{Sanchez-Diaz} {et~al.}(2017){Sanchez-Diaz}, {Rouillard}, {Davies},
  {Lavraud}, {Pinto}, \& {Kilpua}}]{2017ApJ...851...32S}
{Sanchez-Diaz}, E., {Rouillard}, A.~P., {Davies}, J.~A., {et~al.} 2017, \apj,
  851, 32

\bibitem[{Savani {et~al.}(2010)Savani, Owens, Rouillard, Forsyth, \&
  Davies}]{savani2010observational}
Savani, N.~P., Owens, M.~J., Rouillard, A., Forsyth, R., \& Davies, J. 2010,
  The Astrophysical Journal Letters, 714, L128

\bibitem[{{Schwenn} {et~al.}(2006){Schwenn}, {Raymond}, {Alexander},
  {Ciaravella}, {Gopalswamy}, {Howard}, {Hudson}, {Kaufmann}, {Klassen},
  {Maia}, {Munoz-Martinez}, {Pick}, {Reiner}, {Srivastava}, {Tripathi},
  {Vourlidas}, {Wang}, \& {Zhang}}]{2006cme..book..127S}
{Schwenn}, R., {Raymond}, J.~C., {Alexander}, D., {et~al.} 2006, {Coronal
  Observations of CMEs}, Vol.~21, 127

\bibitem[{Sheeley {et~al.}(1997)Sheeley, Wang, Hawley, Brueckner, \&
  Biesecker}]{1997Measurements}
Sheeley, R., Wang, Y.~M., Hawley, H., Brueckner, G.~E., \& Biesecker, A. 1997,
  Astrophysical Journal, 484, 472

\bibitem[{Sheeley~Jr {et~al.}(1999)Sheeley~Jr, Walters, Wang, \&
  Howard}]{sheeley1999continuous}
Sheeley~Jr, N., Walters, J., Wang, Y.-M., \& Howard, R. 1999, Journal of
  Geophysical Research: Space Physics, 104, 24739

\bibitem[{Thernisien(2011)}]{thernisien2011implementation}
Thernisien, A. 2011, The Astrophysical Journal Supplement Series, 194, 33

\bibitem[{{Thernisien} {et~al.}(2006){Thernisien}, {Howard}, \&
  {Vourlidas}}]{2006ApJ...652..763T}
{Thernisien}, A.~F.~R., {Howard}, R.~A., \& {Vourlidas}, A. 2006, \apj, 652,
  763

\bibitem[{{Volpes} \& {Bothmer}(2016)}]{2016AIPC.1714c0003V}
{Volpes}, L. \& {Bothmer}, V. 2016, in American Institute of Physics Conference
  Series, Vol. 1714, Space Plasma Physics, 030003

\bibitem[{{Vourlidas} \& {Howard}(2006)}]{2006ApJ...642.1216V}
{Vourlidas}, A. \& {Howard}, R.~A. 2006, \apj, 642, 1216

\bibitem[{{Vourlidas} {et~al.}(2016){Vourlidas}, {Howard}, {Plunkett},
  {Korendyke}, {Thernisien}, {Wang}, {Rich}, {Carter}, {Chua}, {Socker},
  {Linton}, {Morrill}, {Lynch}, {Thurn}, {Van Duyne}, {Hagood}, {Clifford},
  {Grey}, {Velli}, {Liewer}, {Hall}, {DeJong}, {Mikic}, {Rochus}, {Mazy},
  {Bothmer}, \& {Rodmann}}]{2016SSRv..204...83V}
{Vourlidas}, A., {Howard}, R.~A., {Plunkett}, S.~P., {et~al.} 2016, \ssr, 204,
  83

\bibitem[{{Vr{\v{s}}nak} {et~al.}(2019){Vr{\v{s}}nak}, {Amerstorfer},
  {Dumbovi{\'c}}, {Leitner}, {Veronig}, {Temmer}, {M{\"o}stl}, {Amerstorfer},
  {Farrugia}, \& {Galvin}}]{2019ApJ...877...77V}
{Vr{\v{s}}nak}, B., {Amerstorfer}, T., {Dumbovi{\'c}}, M., {et~al.} 2019, \apj,
  877, 77

\bibitem[{Wang {et~al.}(2005)Wang, Du, \& Richardson}]{wang2005characteristics}
Wang, C., Du, D., \& Richardson, J. 2005, Journal of Geophysical Research:
  Space Physics, 110

\bibitem[{{Wang} {et~al.}(2020){Wang}, {Ji}, {Wang}, {Xia}, {Shen}, {Guo},
  {Zhang}, {Huang}, {Liu}, {Li}, {Liu}, {Wang}, \&
  {Wang}}]{2020ScChE..63.1699W}
{Wang}, Y., {Ji}, H., {Wang}, Y., {et~al.} 2020, Science in China E:
  Technological Sciences, 63, 1699

\bibitem[{{Wang} {et~al.}(2015){Wang}, {Zhou}, {Shen}, {Liu}, \&
  {Wang}}]{2015JGRA..120.1543W}
{Wang}, Y., {Zhou}, Z., {Shen}, C., {Liu}, R., \& {Wang}, S. 2015, Journal of
  Geophysical Research (Space Physics), 120, 1543

\bibitem[{{Wang} \& {Sheeley}(1990)}]{1990ApJ...355..726W}
{Wang}, Y.~M. \& {Sheeley}, N.~R., J. 1990, \apj, 355, 726

\bibitem[{{Webb} \& {Howard}(2012)}]{2012LRSP....9....3W}
{Webb}, D.~F. \& {Howard}, T.~A. 2012, Living Reviews in Solar Physics, 9, 3

\bibitem[{Wood {et~al.}(2016)Wood, Howard, \& Linton}]{wood2016imaging}
Wood, B.~E., Howard, R.~A., \& Linton, M.~G. 2016, The Astrophysical Journal,
  816, 67

\bibitem[{{Wood} {et~al.}(2011){Wood}, {Wu}, {Howard}, {Socker}, \&
  {Rouillard}}]{2011ApJ...729...70W}
{Wood}, B.~E., {Wu}, C.~C., {Howard}, R.~A., {Socker}, D.~G., \& {Rouillard},
  A.~P. 2011, \apj, 729, 70

\bibitem[{{Xie} {et~al.}(2012){Xie}, {Odstrcil}, {Mays}, {St. Cyr},
  {Gopalswamy}, \& {Cremades}}]{2012JGRA..117.4105X}
{Xie}, H., {Odstrcil}, D., {Mays}, L., {et~al.} 2012, Journal of Geophysical
  Research (Space Physics), 117, A04105

\bibitem[{Ying {et~al.}(2019)Ying, Bemporad, Giordano, Pagano, Feng, Lu, Li, \&
  Gan}]{ying2019first}
Ying, B., Bemporad, A., Giordano, S., {et~al.} 2019, The Astrophysical Journal,
  880, 41

\end{thebibliography}

  \begin{table*}
  \caption{Estimated radial velocity ($v_r$) of the CME ejecta leading edge on 3 April 2010}             
  \label{table:1}      
  \centering                          
  \begin{tabular}{l l r r r r r r}        
  \hline\hline                 
Method\tablefootmark{a} & spacecraft & imager &$\lambda$\tablefootmark{b} [$^\circ$] &$\Phi$\tablefootmark{b} [$^\circ$] & \multicolumn{3}{c}{$v_r$ [km/s]}\\    &  &  &  &  & at 15:29 UT ($D \tablefootmark{b}\approx 30R_{\odot}$) & at 22:49 UT ($D\tablefootmark{b} \approx 60R_{\odot}$) &mean $v_r$\tablefootmark{c}\\
  \hline
   MCT              & STEREO-A/B   & HI1 & $\approx0$ & 7          & $877\pm 33$  & $834\pm29$  & $855\pm33$  \\      
  \hline
  \multicolumn{6}{l}{\it Results of other methods in other publications}\\
   FP$^1$           & STEREO-A     & HI1     & 0 & 9          & $950\pm150$ \tablefootmark{d} &$950\pm150$ \tablefootmark{d}& $-$         \\
   FP$^2$           & STEREO-A     & HI1/HI2 & 0 & $3\pm4$    & $800\pm100$ \tablefootmark{d} &$850\pm100$ \tablefootmark{d}& $829\pm122$ \\
   FP$^3$           & STEREO-A     & HI1     & 0 & $5\pm2$    & $\approx800$  \tablefootmark{d}  &$\approx 700$ \tablefootmark{d}& $-$ \\
   HM$^1$           & STEREO-A     & HI1     & 0 & $-5$       & $950\pm150$ \tablefootmark{d} &$950\pm150$ \tablefootmark{d}& $-$         \\
   HM$^2$           & STEREO-A     & HI1/HI2 & 0 & $-25\pm10$ & $750\pm100$ \tablefootmark{d} &$950\pm100$ \tablefootmark{d}& $854\pm100$ \\
   HM$^3$           & STEREO-A     & HI1     & 0 & $6\pm2$    & $\approx800$ \tablefootmark{d}   &$\approx 900$ \tablefootmark{d}& $-$ \\
   FP/HM$^4$        & STEREO-A     & HI1     & 0 & $-8$       & $1100\pm100$ \tablefootmark{d}&$950\pm100$ \tablefootmark{d}& $-$         \\
   GT$^5$           & STEREO-A/B   & HI1     & 0 & 10         & $900\pm150$ \tablefootmark{d} &$750\pm200$ \tablefootmark{d}& $-$         \\
   SSEF$^6$         & STEREO-A     & HI1/HI2 & 0 & $-19$      & $-$          & $-$         & 915       \\
  \hline
  \multicolumn{6}{l}{\it Results of  HELCATS$^{e,7,8}$}\\
   FP            & STEREO-A     & HI1/HI2 & $-2$ & 0       & $-$          & $-$         & $889\pm17$ \\
   FP            & STEREO-B     & HI1/HI2 & $-5$ & 34      & $-$          & $-$         & $1149\pm106$ \\
   HM            & STEREO-A     & HI1/HI2 & $-2$ & $-20$   & $-$          & $-$         & $962\pm24$ \\
   HM            & STEREO-B     & HI1/HI2 & $-3$ & 69      & $-$          & $-$         & $1368\pm213$ \\
   SSEF          & STEREO-A     & HI1/HI2 & $-2$ & $-10$   & $-$          & $-$         & $927\pm20$ \\
   SSEF          & STEREO-B     & HI1/HI2 & $-5$ & 51      & $-$          & $-$         & $1248\pm150$ \\
  \hline                                 

  \end{tabular}
  \tablefoot{\\
  \tablefoottext{a}{The full names of the abbreviations of the methods are listed below, MCT: maximum correlation-coefficient localization and cross-correlation tracking technique, FP: fixed-$\Phi$ fitting method, HM: harmonic mean fitting method, GT: geometric rriangulation method, SSEF: self-similar expansion fitting method.}\\
  \tablefoottext{b}{$\lambda$, $\Phi$ and $D$ are the latitude, longitude, and heliocentric distance, respectively, in HEE coordinates.}\\
  \tablefoottext{c}{The mean velocity of MCT method is calculated for a heliocentric distance ($D$) between $30 R_\odot$ and $60 R_\odot$, while for the other methods, the mean velocity is derived by fitting the whole trace of the leading edge in HI images.}\\
  \tablefoottext{d}{The velocity here is roughly estimated from the velocity images in corresponding papers.}\\
  \tablefoottext{e}{Data is obtained from the Heliospheric Imager CME catalog and the CME kinematics catalog of HEliospheric Cataloging, Analysis and Techiques Service (HELCATS) (https://www.helcats-fp7.eu/index.html).}
  \tablebib{(1)~\citet{2010GeoRL..3724103M};
(2) \citet{2012SoPh..276..293R}; (3) \citet{2016AIPC.1714c0003V}; (4) \citet{2012JGRA..117.4105X}; (5) \citet{2011ApJ...734...84L};
(6) \citet{2014ApJ...787..119M}; (7) \citet{2013ApJ...777..167D}; (8) \citet{2019SoPh..294...57B}.}
  }
   \label{Tab1}
  \end{table*}
\end{document}